\documentstyle[twoside,fleqn,espcrc2]{article}
\newcommand{\AmS}{{\protect\the\textfont2
  A\kern-.1667em\lower.5ex\hbox{M}\kern-.125emS}}
\begin{document}

\title{Eigenmodes of the Dirac Operator 
and Chiral Properties of QCD with Sea Quarks}

\author{Xiang-Qian Luo 
\address{Department of Physics,
       Zhongshan University, Guangzhou 510275, 
       China}
 }

\begin{abstract}
I describe a mechanism to understand 
the relation between chiral-symmetry breaking 
and eigenmodes of the Dirac operator 
in lattice QCD with Kogut-Susskind sea quarks. 
It can be shown that if chiral symmetry
is spontaneously broken, the eigenvalues $\lambda_i$
should behave as $z(i)/V$ for large volume $V$, where
$z(i)$ is the i-th zero of the Bessel function.
With neither chiral nor $\lambda$ extrapolation, 
one can precisely calculate the chiral condensate using 
only a small set of eigenvalues. Therefore,
it is economical and 
free of systematic uncertainties. I present the first QCD
data to support this mechanism and
encourage the lattice community to test and use
it.
\end{abstract}

\maketitle


One can not completely understand physics of hadrons 
without understanding the QCD vacuum, 
which main properties 
are confinement and chiral-symmetry breaking.

The chiral order parameter is $\langle {\bar \psi} \psi \rangle$
in the chiral limit. 
Suppose $\langle {\bar \chi} \chi (m,V) \rangle$ is the one
calculated at bare mass $m$ and lattice volume $V$.
Since  $\langle {\bar \chi} \chi (0,V) \rangle =0$, 
conventionally one has to assume 
\begin{eqnarray}
\langle {\bar \psi} \psi \rangle 
= \lim_{m \to 0} \lim_{V \to \infty} 
\langle {\bar \chi} \chi (m,V) \rangle.
\label{ps}
\end{eqnarray}
Practically,
most people do extrapolation only on finite $V$ and very small
set of $m$ by means of some unjustified fitting function
(linear, quadratic, polynomial or logarithm ...).
Unfortunately, such an process is arbitrary and
sometimes gives wrong results. Here we provide 
two known evidences.

\noindent
(i) QED in 4 dimensions. Non-compact lattice QED 
experiences second order 
chiral phase transition at finite
some bare coupling constant $g$, where the chiral
condensate vanishes.
The critical coupling determined by
naive $m$ extrapolation of $\langle {\bar \chi} \chi (m,V) \rangle$
is not necessary correct.
For detailed discussions and current status, 
see \cite{QED4}. 

\noindent
(ii) The one-flavor massless Schwinger model. 
In the continuum, as is well known,
$\langle {\bar \psi} \psi \rangle_{cont}
= 0.16 e$, where $e$ is the electric charge.
As indicated in Ref. \cite{QED2}, the exact result
can not be reached  at finite bare coupling $g=ae$
by the any extrapolation mentioned above.

In Ref. \cite{PDF}, we proposed a new mechanism
to get reliable quantitative information 
on chiral-symmetry breaking,
based on the computation of the probability distribution 
function of $\langle {\bar \psi} \psi \rangle$:
\begin{eqnarray}
P(c) = \lim_{N \to \infty} \langle \delta (c - 
{1 \over N}\sum_x {\bar \psi}(x) \psi(x) )
\rangle.
\label{pdf}
\end{eqnarray}
Here $N$ is the degrees of freedom,
being proportional to $V$.
$P(c)$ is the probability to get the value $c$ 
for the chiral condensate.
If there is exact chiral symmetry in 
the ground state, $P(c)=\delta(c)$.
If chiral symmetry is spontaneously broken,
$P(c)$ will be a more complex function.

For Kogut-Susskind fermions,
we derive several relations between the chiral condensate
and the eigenmodes of the Dirac operator. 

\noindent
(a) The first formula is
\begin{eqnarray}
\langle {\bar \psi} \psi \rangle 
= \lim_{m \to 0} \lim_{V \to \infty} 
{1 \over N} \langle \sum_i^{N} 
{2m \over \lambda_i^2 +m^2} \rangle 
\label{ps1}
\end{eqnarray}
where $\lambda_i$ is the i-th positive eigenvalue 
of the massless Dirac matrix.
This is the formula
many people are currently using.
It requires $m$ extrapolation,
and may not produce reliable result.

\noindent
(b) The second formula is
\begin{eqnarray}
\langle {\bar \psi} \psi \rangle 
= \lim_{\lambda \to 0} \lim_{V \to \infty} 
 {\pi \over N} \langle \rho(\lambda) \rangle,
\label{ps2}
\end{eqnarray}
where $\rho(\lambda)$ is the eigenmode density.
This formula is also known. Since $\rho(0)=0$ at
finite $V$, and it requires $\lambda$ 
extrapolation, which is again arbitrary. 
 
\noindent
(c) The third formula \cite{PDF} is
\begin{eqnarray}
\langle {\bar \psi} \psi \rangle 
= \lim_{V \to \infty} 
\sqrt {{4 \over N^2} \langle \sum_{i=1}^N {1\over \lambda_i^2} \rangle},
\label{ps3}
\end{eqnarray}
For its derivation by 
other methods, see Refs. \cite{LS,Ver,DMW}.
The advantage over the first two formulae is that
no $\lambda$ extrapolation is necessary. 
The disadvantage is that all the eigenvalues 
have to be calculated. When the lattice volume is
large, the computational task is huge and not so
feasible. Indeed, we have checked that it requires
a large lattice volume for getting consistent results.

\noindent
(d) The fourth formula is
\begin{eqnarray}
<{\bar \psi} \psi> =C(i)
= \lim_{N \to \infty} {z(i) \over N} 
\langle {1 \over \lambda_i} \rangle.
\label{ps4}
\end{eqnarray}
Here $z(i)$ is the i-th zero of the Bessel function 
$J_0$.
This is new and was originally 
derived in Ref. \cite{PDF}.
Neither $m$ nor $\lambda$ extrapolation is necessary.
The advantage over the third formula is that
only a few smallest eigenvalues are needed 
for this calculation. 
The only thing is the finite size analysis
which every formula requires to do.

The relation between eigenmodes 
and chiral-symmetry breaking is clear:
if chiral symmetry
is spontaneously broken, $\lambda_i$
should behave as $z(i)/V$ for large $V$.
We have numerically 
checked \cite{PDF} the reliability of Eq. (\ref{ps4})
in the Schwinger model, where
the exact result was reproduced precisely.
To our knowledge, this is one of the best results 
for the lattice Schwinger model!

The most interesting application is QCD.
Here I would like to present the first data for QCD
with various numbers of quark flavors $N_f$.
In Ref. \cite{Luo}, the configurations
were generated and eigenvalues of the fermionic
matrix were calculated.
With these, it is very easy to evaluate the
chiral condensate and compare different methods.

\begin{figure}[htb]
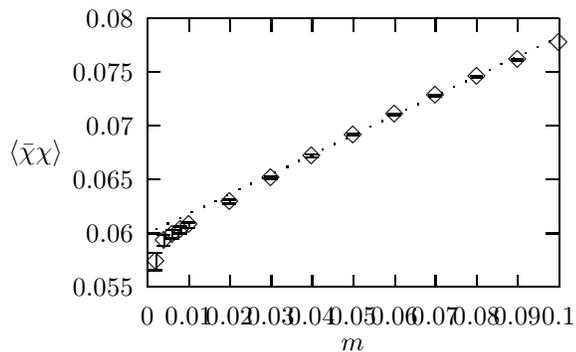

%\fpsxsize=6.2cm
\hspace{6mm}
%\def\fpsangle{270}
%insert the figure here
% GNUPLOT: LaTeX picture
\setlength{\unitlength}{0.240900pt}
\ifx\plotpoint\undefined\newsavebox{\plotpoint}\fi
\font\gnuplot=cmr10 at 10pt
\sbox{\plotpoint}{\rule[-0.200pt]{0.400pt}{0.400pt}}%
\put(220.0,113.0){\rule[-0.200pt]{0.400pt}{101.660pt}}
\put(220.0,113.0){\rule[-0.200pt]{4.818pt}{0.400pt}}
\put(198,113){\makebox(0,0)[r]{0.055}}
\put(846.0,113.0){\rule[-0.200pt]{4.818pt}{0.400pt}}
\put(220.0,197.0){\rule[-0.200pt]{4.818pt}{0.400pt}}
\put(198,197){\makebox(0,0)[r]{0.06}}
\put(846.0,197.0){\rule[-0.200pt]{4.818pt}{0.400pt}}
\put(220.0,282.0){\rule[-0.200pt]{4.818pt}{0.400pt}}
\put(198,282){\makebox(0,0)[r]{0.065}}
\put(846.0,282.0){\rule[-0.200pt]{4.818pt}{0.400pt}}
\put(220.0,366.0){\rule[-0.200pt]{4.818pt}{0.400pt}}
\put(198,366){\makebox(0,0)[r]{0.07}}
\put(846.0,366.0){\rule[-0.200pt]{4.818pt}{0.400pt}}
\put(220.0,451.0){\rule[-0.200pt]{4.818pt}{0.400pt}}
\put(198,451){\makebox(0,0)[r]{0.075}}
\put(846.0,451.0){\rule[-0.200pt]{4.818pt}{0.400pt}}
\put(220.0,535.0){\rule[-0.200pt]{4.818pt}{0.400pt}}
\put(198,535){\makebox(0,0)[r]{0.08}}
\put(846.0,535.0){\rule[-0.200pt]{4.818pt}{0.400pt}}
\put(220.0,113.0){\rule[-0.200pt]{0.400pt}{4.818pt}}
\put(220,68){\makebox(0,0){0}}
\put(220.0,515.0){\rule[-0.200pt]{0.400pt}{4.818pt}}
\put(285.0,113.0){\rule[-0.200pt]{0.400pt}{4.818pt}}
\put(285,68){\makebox(0,0){0.01}}
\put(285.0,515.0){\rule[-0.200pt]{0.400pt}{4.818pt}}
\put(349.0,113.0){\rule[-0.200pt]{0.400pt}{4.818pt}}
\put(349,68){\makebox(0,0){0.02}}
\put(349.0,515.0){\rule[-0.200pt]{0.400pt}{4.818pt}}
\put(414.0,113.0){\rule[-0.200pt]{0.400pt}{4.818pt}}
\put(414,68){\makebox(0,0){0.03}}
\put(414.0,515.0){\rule[-0.200pt]{0.400pt}{4.818pt}}
\put(478.0,113.0){\rule[-0.200pt]{0.400pt}{4.818pt}}
\put(478,68){\makebox(0,0){0.04}}
\put(478.0,515.0){\rule[-0.200pt]{0.400pt}{4.818pt}}
\put(543.0,113.0){\rule[-0.200pt]{0.400pt}{4.818pt}}
\put(543,68){\makebox(0,0){0.05}}
\put(543.0,515.0){\rule[-0.200pt]{0.400pt}{4.818pt}}
\put(608.0,113.0){\rule[-0.200pt]{0.400pt}{4.818pt}}
\put(608,68){\makebox(0,0){0.06}}
\put(608.0,515.0){\rule[-0.200pt]{0.400pt}{4.818pt}}
\put(672.0,113.0){\rule[-0.200pt]{0.400pt}{4.818pt}}
\put(672,68){\makebox(0,0){0.07}}
\put(672.0,515.0){\rule[-0.200pt]{0.400pt}{4.818pt}}
\put(737.0,113.0){\rule[-0.200pt]{0.400pt}{4.818pt}}
\put(737,68){\makebox(0,0){0.08}}
\put(737.0,515.0){\rule[-0.200pt]{0.400pt}{4.818pt}}
\put(801.0,113.0){\rule[-0.200pt]{0.400pt}{4.818pt}}
\put(801,68){\makebox(0,0){0.09}}
\put(801.0,515.0){\rule[-0.200pt]{0.400pt}{4.818pt}}
\put(866.0,113.0){\rule[-0.200pt]{0.400pt}{4.818pt}}
\put(866,68){\makebox(0,0){0.1}}
\put(866.0,515.0){\rule[-0.200pt]{0.400pt}{4.818pt}}
\put(220.0,113.0){\rule[-0.200pt]{155.621pt}{0.400pt}}
\put(866.0,113.0){\rule[-0.200pt]{0.400pt}{101.660pt}}
\put(220.0,535.0){\rule[-0.200pt]{155.621pt}{0.400pt}}
\put(45,324){\makebox(0,0){$\langle \bar{\chi} \chi \rangle$}}
\put(543,23){\makebox(0,0){$m$}}
\put(220.0,113.0){\rule[-0.200pt]{0.400pt}{101.660pt}}
\put(233,153){\raisebox{-.8pt}{\makebox(0,0){$\Diamond$}}}
\put(246,186){\raisebox{-.8pt}{\makebox(0,0){$\Diamond$}}}
\put(259,195){\raisebox{-.8pt}{\makebox(0,0){$\Diamond$}}}
\put(272,203){\raisebox{-.8pt}{\makebox(0,0){$\Diamond$}}}
\put(285,210){\raisebox{-.8pt}{\makebox(0,0){$\Diamond$}}}
\put(349,247){\raisebox{-.8pt}{\makebox(0,0){$\Diamond$}}}
\put(414,284){\raisebox{-.8pt}{\makebox(0,0){$\Diamond$}}}
\put(478,319){\raisebox{-.8pt}{\makebox(0,0){$\Diamond$}}}
\put(543,352){\raisebox{-.8pt}{\makebox(0,0){$\Diamond$}}}
\put(608,384){\raisebox{-.8pt}{\makebox(0,0){$\Diamond$}}}
\put(672,414){\raisebox{-.8pt}{\makebox(0,0){$\Diamond$}}}
\put(737,443){\raisebox{-.8pt}{\makebox(0,0){$\Diamond$}}}
\put(801,470){\raisebox{-.8pt}{\makebox(0,0){$\Diamond$}}}
\put(866,496){\raisebox{-.8pt}{\makebox(0,0){$\Diamond$}}}
\put(233.0,139.0){\rule[-0.200pt]{0.400pt}{6.745pt}}
\put(223.0,139.0){\rule[-0.200pt]{4.818pt}{0.400pt}}
\put(223.0,167.0){\rule[-0.200pt]{4.818pt}{0.400pt}}
\put(246.0,178.0){\rule[-0.200pt]{0.400pt}{4.095pt}}
\put(236.0,178.0){\rule[-0.200pt]{4.818pt}{0.400pt}}
\put(236.0,195.0){\rule[-0.200pt]{4.818pt}{0.400pt}}
\put(259.0,189.0){\rule[-0.200pt]{0.400pt}{3.132pt}}
\put(249.0,189.0){\rule[-0.200pt]{4.818pt}{0.400pt}}
\put(249.0,202.0){\rule[-0.200pt]{4.818pt}{0.400pt}}
\put(272.0,197.0){\rule[-0.200pt]{0.400pt}{2.650pt}}
\put(262.0,197.0){\rule[-0.200pt]{4.818pt}{0.400pt}}
\put(262.0,208.0){\rule[-0.200pt]{4.818pt}{0.400pt}}
\put(285.0,206.0){\rule[-0.200pt]{0.400pt}{1.927pt}}
\put(275.0,206.0){\rule[-0.200pt]{4.818pt}{0.400pt}}
\put(275.0,214.0){\rule[-0.200pt]{4.818pt}{0.400pt}}
\put(349.0,244.0){\rule[-0.200pt]{0.400pt}{1.445pt}}
\put(339.0,244.0){\rule[-0.200pt]{4.818pt}{0.400pt}}
\put(339.0,250.0){\rule[-0.200pt]{4.818pt}{0.400pt}}
\put(414.0,282.0){\rule[-0.200pt]{0.400pt}{0.964pt}}
\put(404.0,282.0){\rule[-0.200pt]{4.818pt}{0.400pt}}
\put(404.0,286.0){\rule[-0.200pt]{4.818pt}{0.400pt}}
\put(478.0,317.0){\rule[-0.200pt]{0.400pt}{0.964pt}}
\put(468.0,317.0){\rule[-0.200pt]{4.818pt}{0.400pt}}
\put(468.0,321.0){\rule[-0.200pt]{4.818pt}{0.400pt}}
\put(543.0,351.0){\rule[-0.200pt]{0.400pt}{0.723pt}}
\put(533.0,351.0){\rule[-0.200pt]{4.818pt}{0.400pt}}
\put(533.0,354.0){\rule[-0.200pt]{4.818pt}{0.400pt}}
\put(608.0,382.0){\rule[-0.200pt]{0.400pt}{0.723pt}}
\put(598.0,382.0){\rule[-0.200pt]{4.818pt}{0.400pt}}
\put(598.0,385.0){\rule[-0.200pt]{4.818pt}{0.400pt}}
\put(672.0,412.0){\rule[-0.200pt]{0.400pt}{0.723pt}}
\put(662.0,412.0){\rule[-0.200pt]{4.818pt}{0.400pt}}
\put(662.0,415.0){\rule[-0.200pt]{4.818pt}{0.400pt}}
\put(737.0,441.0){\rule[-0.200pt]{0.400pt}{0.723pt}}
\put(727.0,441.0){\rule[-0.200pt]{4.818pt}{0.400pt}}
\put(727.0,444.0){\rule[-0.200pt]{4.818pt}{0.400pt}}
\put(801.0,469.0){\rule[-0.200pt]{0.400pt}{0.482pt}}
\put(791.0,469.0){\rule[-0.200pt]{4.818pt}{0.400pt}}
\put(791.0,471.0){\rule[-0.200pt]{4.818pt}{0.400pt}}
\put(233,204){\usebox{\plotpoint}}
\put(233.00,204.00){\usebox{\plotpoint}}
\multiput(239,207)(19.077,8.176){0}{\usebox{\plotpoint}}
\put(251.75,212.88){\usebox{\plotpoint}}
\multiput(252,213)(18.564,9.282){0}{\usebox{\plotpoint}}
\multiput(258,216)(19.077,8.176){0}{\usebox{\plotpoint}}
\put(270.51,221.75){\usebox{\plotpoint}}
\multiput(271,222)(19.077,8.176){0}{\usebox{\plotpoint}}
\multiput(278,225)(18.564,9.282){0}{\usebox{\plotpoint}}
\put(289.26,230.63){\usebox{\plotpoint}}
\multiput(290,231)(19.077,8.176){0}{\usebox{\plotpoint}}
\multiput(297,234)(18.564,9.282){0}{\usebox{\plotpoint}}
\put(308.15,239.21){\usebox{\plotpoint}}
\multiput(310,240)(18.564,9.282){0}{\usebox{\plotpoint}}
\multiput(316,243)(18.564,9.282){0}{\usebox{\plotpoint}}
\put(326.89,248.10){\usebox{\plotpoint}}
\multiput(329,249)(17.270,11.513){0}{\usebox{\plotpoint}}
\multiput(335,253)(19.077,8.176){0}{\usebox{\plotpoint}}
\put(345.25,257.63){\usebox{\plotpoint}}
\multiput(348,259)(18.564,9.282){0}{\usebox{\plotpoint}}
\multiput(354,262)(19.077,8.176){0}{\usebox{\plotpoint}}
\put(364.01,266.50){\usebox{\plotpoint}}
\multiput(367,268)(19.077,8.176){0}{\usebox{\plotpoint}}
\multiput(374,271)(18.564,9.282){0}{\usebox{\plotpoint}}
\put(382.76,275.38){\usebox{\plotpoint}}
\multiput(386,277)(19.077,8.176){0}{\usebox{\plotpoint}}
\multiput(393,280)(18.564,9.282){0}{\usebox{\plotpoint}}
\put(401.58,284.11){\usebox{\plotpoint}}
\multiput(406,286)(18.564,9.282){0}{\usebox{\plotpoint}}
\multiput(412,289)(18.564,9.282){0}{\usebox{\plotpoint}}
\put(420.33,293.00){\usebox{\plotpoint}}
\multiput(425,295)(18.564,9.282){0}{\usebox{\plotpoint}}
\multiput(431,298)(19.077,8.176){0}{\usebox{\plotpoint}}
\put(439.12,301.75){\usebox{\plotpoint}}
\multiput(444,305)(18.564,9.282){0}{\usebox{\plotpoint}}
\multiput(450,308)(19.077,8.176){0}{\usebox{\plotpoint}}
\put(457.51,311.25){\usebox{\plotpoint}}
\multiput(463,314)(19.077,8.176){0}{\usebox{\plotpoint}}
\multiput(470,317)(18.564,9.282){0}{\usebox{\plotpoint}}
\put(476.26,320.13){\usebox{\plotpoint}}
\multiput(482,323)(19.077,8.176){0}{\usebox{\plotpoint}}
\multiput(489,326)(18.564,9.282){0}{\usebox{\plotpoint}}
\put(495.01,329.01){\usebox{\plotpoint}}
\multiput(501,332)(19.077,8.176){0}{\usebox{\plotpoint}}
\put(513.76,337.88){\usebox{\plotpoint}}
\multiput(514,338)(19.077,8.176){0}{\usebox{\plotpoint}}
\multiput(521,341)(18.564,9.282){0}{\usebox{\plotpoint}}
\put(532.52,346.76){\usebox{\plotpoint}}
\multiput(533,347)(19.077,8.176){0}{\usebox{\plotpoint}}
\multiput(540,350)(18.564,9.282){0}{\usebox{\plotpoint}}
\put(551.12,355.92){\usebox{\plotpoint}}
\multiput(553,357)(18.564,9.282){0}{\usebox{\plotpoint}}
\multiput(559,360)(18.564,9.282){0}{\usebox{\plotpoint}}
\put(569.75,365.04){\usebox{\plotpoint}}
\multiput(572,366)(18.564,9.282){0}{\usebox{\plotpoint}}
\multiput(578,369)(19.077,8.176){0}{\usebox{\plotpoint}}
\put(588.56,373.78){\usebox{\plotpoint}}
\multiput(591,375)(18.564,9.282){0}{\usebox{\plotpoint}}
\multiput(597,378)(19.077,8.176){0}{\usebox{\plotpoint}}
\put(607.32,382.66){\usebox{\plotpoint}}
\multiput(610,384)(19.077,8.176){0}{\usebox{\plotpoint}}
\multiput(617,387)(18.564,9.282){0}{\usebox{\plotpoint}}
\put(626.07,391.53){\usebox{\plotpoint}}
\multiput(629,393)(19.077,8.176){0}{\usebox{\plotpoint}}
\multiput(636,396)(18.564,9.282){0}{\usebox{\plotpoint}}
\put(644.90,400.24){\usebox{\plotpoint}}
\multiput(649,402)(17.270,11.513){0}{\usebox{\plotpoint}}
\multiput(655,406)(18.564,9.282){0}{\usebox{\plotpoint}}
\put(663.18,409.94){\usebox{\plotpoint}}
\multiput(668,412)(18.564,9.282){0}{\usebox{\plotpoint}}
\multiput(674,415)(19.077,8.176){0}{\usebox{\plotpoint}}
\put(682.06,418.53){\usebox{\plotpoint}}
\multiput(687,421)(18.564,9.282){0}{\usebox{\plotpoint}}
\multiput(693,424)(19.077,8.176){0}{\usebox{\plotpoint}}
\put(700.82,427.41){\usebox{\plotpoint}}
\multiput(706,430)(19.077,8.176){0}{\usebox{\plotpoint}}
\multiput(713,433)(18.564,9.282){0}{\usebox{\plotpoint}}
\put(719.57,436.28){\usebox{\plotpoint}}
\multiput(725,439)(19.077,8.176){0}{\usebox{\plotpoint}}
\multiput(732,442)(18.564,9.282){0}{\usebox{\plotpoint}}
\put(738.32,445.16){\usebox{\plotpoint}}
\multiput(744,448)(19.077,8.176){0}{\usebox{\plotpoint}}
\multiput(751,451)(18.564,9.282){0}{\usebox{\plotpoint}}
\put(757.07,454.04){\usebox{\plotpoint}}
\multiput(764,458)(18.564,9.282){0}{\usebox{\plotpoint}}
\put(775.43,463.71){\usebox{\plotpoint}}
\multiput(776,464)(19.077,8.176){0}{\usebox{\plotpoint}}
\multiput(783,467)(18.564,9.282){0}{\usebox{\plotpoint}}
\put(794.32,472.28){\usebox{\plotpoint}}
\multiput(796,473)(18.564,9.282){0}{\usebox{\plotpoint}}
\multiput(802,476)(18.564,9.282){0}{\usebox{\plotpoint}}
\put(813.07,481.17){\usebox{\plotpoint}}
\multiput(815,482)(18.564,9.282){0}{\usebox{\plotpoint}}
\multiput(821,485)(19.077,8.176){0}{\usebox{\plotpoint}}
\put(831.87,489.94){\usebox{\plotpoint}}
\multiput(834,491)(18.564,9.282){0}{\usebox{\plotpoint}}
\multiput(840,494)(19.077,8.176){0}{\usebox{\plotpoint}}
\put(850.63,498.81){\usebox{\plotpoint}}
\multiput(853,500)(19.077,8.176){0}{\usebox{\plotpoint}}
\multiput(860,503)(18.564,9.282){0}{\usebox{\plotpoint}}
\put(866,506){\usebox{\plotpoint}}
\vspace{-10mm}
\caption{$\langle \bar{\chi} \chi (m,V) \rangle$
for $N_f=0$, $\beta=5.5088$ and $V=6^4$
using Eq. (3).}
\label{fig1}
\end{figure}

Figure \ref{fig1} shows the quenched result for
$\beta=5.5088$ and $V=6^4$
using the first formula: Eq. (\ref{ps1}). 
Only a small set of data can be fitted by a linear function.
If one fits only the data 
for $m \in [0.04, 0.08]$ to the chiral limit, 
the result is 0.060. But outside this range,
either for larger or smaller $m$, this slope
changes and global fitting function can not be easily found. For smaller $m$, 
the change is very rapid. We attribute it to
finite size effects. 

Figure \ref{fig2} shows the quenched result for
$\beta=5.5088$ and $V=6^4$, but
using the fourth formula: Eq. (\ref{ps4}).
Even on such a small lattice, there is a nice
plateau for $i \in [10,25]$, from which
one can reliably get $C(i)=0.06$. 
No fitting function is necessary. 
Of course, detailed finite size analysis 
remains to be done.

Figures \ref{fig3}, \ref{fig4}, \ref{fig5}, \ref{fig6} 
are the results for $N_f=1, 2, 3, 4$.
All these satisfy well Eq. (\ref{ps4}).

In conclusion,
I have shown a reliable method for investigating 
the chiral properties and 
obtaining the chiral condensate from
only a small set of $\lambda$,
without $m$ or $\lambda$ extrapolation. 
This method was previously tested in the Schwinger model.
In this paper I have presented the first data for QCD.
Therefore, I encourage other people 
in the lattice community to test and use Eq. (\ref{ps4}).

This work was supported by the
National Natural Science Fund for Distinguished Young Scholars,
supplemented by the
National Natural Science Foundation, 
fund for international cooperation and exchange,
the Ministry of Education, 
the Zhongshan University Administrations
and Hong Kong Foundation of
the Zhongshan University Advanced Research Center. 
I also
thank the Lattice 98 organizers
for additional support and assistance.
Discussions with V.Azcoiti, T.Chiu, G.DiCarlo, 
A.Galante, A.Grillo, 
H. Jirari, H. Kr\"oger, and V. Laliena are appreciated.

\begin{figure}[htb]
%\fpsxsize=6.2cm
\hspace{6mm}
%\def\fpsangle{270}
%insert the figure here
% GNUPLOT: LaTeX picture
\setlength{\unitlength}{0.240900pt}
\ifx\plotpoint\undefined\newsavebox{\plotpoint}\fi
%\begin{picture}(930,558)(0,0)
\font\gnuplot=cmr10 at 10pt
%\gnuplot
\sbox{\plotpoint}{\rule[-0.200pt]{0.400pt}{0.400pt}}%
\put(220.0,113.0){\rule[-0.200pt]{4.818pt}{0.400pt}}
\put(198,113){\makebox(0,0)[r]{0.05}}
\put(846.0,113.0){\rule[-0.200pt]{4.818pt}{0.400pt}}
\put(220.0,155.0){\rule[-0.200pt]{4.818pt}{0.400pt}}
\put(198,155){\makebox(0,0)[r]{0.055}}
\put(846.0,155.0){\rule[-0.200pt]{4.818pt}{0.400pt}}
\put(220.0,197.0){\rule[-0.200pt]{4.818pt}{0.400pt}}
\put(198,197){\makebox(0,0)[r]{0.06}}
\put(846.0,197.0){\rule[-0.200pt]{4.818pt}{0.400pt}}
\put(220.0,240.0){\rule[-0.200pt]{4.818pt}{0.400pt}}
\put(198,240){\makebox(0,0)[r]{0.065}}
\put(846.0,240.0){\rule[-0.200pt]{4.818pt}{0.400pt}}
\put(220.0,282.0){\rule[-0.200pt]{4.818pt}{0.400pt}}
\put(198,282){\makebox(0,0)[r]{0.07}}
\put(846.0,282.0){\rule[-0.200pt]{4.818pt}{0.400pt}}
\put(220.0,324.0){\rule[-0.200pt]{4.818pt}{0.400pt}}
\put(198,324){\makebox(0,0)[r]{0.075}}
\put(846.0,324.0){\rule[-0.200pt]{4.818pt}{0.400pt}}
\put(220.0,366.0){\rule[-0.200pt]{4.818pt}{0.400pt}}
\put(198,366){\makebox(0,0)[r]{0.08}}
\put(846.0,366.0){\rule[-0.200pt]{4.818pt}{0.400pt}}
\put(220.0,408.0){\rule[-0.200pt]{4.818pt}{0.400pt}}
\put(198,408){\makebox(0,0)[r]{0.085}}
\put(846.0,408.0){\rule[-0.200pt]{4.818pt}{0.400pt}}
\put(220.0,451.0){\rule[-0.200pt]{4.818pt}{0.400pt}}
\put(198,451){\makebox(0,0)[r]{0.09}}
\put(846.0,451.0){\rule[-0.200pt]{4.818pt}{0.400pt}}
\put(220.0,493.0){\rule[-0.200pt]{4.818pt}{0.400pt}}
\put(198,493){\makebox(0,0)[r]{0.095}}
\put(846.0,493.0){\rule[-0.200pt]{4.818pt}{0.400pt}}
\put(220.0,535.0){\rule[-0.200pt]{4.818pt}{0.400pt}}
\put(198,535){\makebox(0,0)[r]{0.1}}
\put(846.0,535.0){\rule[-0.200pt]{4.818pt}{0.400pt}}
\put(220.0,113.0){\rule[-0.200pt]{0.400pt}{4.818pt}}
\put(220,68){\makebox(0,0){10}}
\put(220.0,515.0){\rule[-0.200pt]{0.400pt}{4.818pt}}
\put(381.0,113.0){\rule[-0.200pt]{0.400pt}{4.818pt}}
\put(381,68){\makebox(0,0){15}}
\put(381.0,515.0){\rule[-0.200pt]{0.400pt}{4.818pt}}
\put(543.0,113.0){\rule[-0.200pt]{0.400pt}{4.818pt}}
\put(543,68){\makebox(0,0){20}}
\put(543.0,515.0){\rule[-0.200pt]{0.400pt}{4.818pt}}
\put(704.0,113.0){\rule[-0.200pt]{0.400pt}{4.818pt}}
\put(704,68){\makebox(0,0){25}}
\put(704.0,515.0){\rule[-0.200pt]{0.400pt}{4.818pt}}
\put(866.0,113.0){\rule[-0.200pt]{0.400pt}{4.818pt}}
\put(866,68){\makebox(0,0){30}}
\put(866.0,515.0){\rule[-0.200pt]{0.400pt}{4.818pt}}
\put(220.0,113.0){\rule[-0.200pt]{155.621pt}{0.400pt}}
\put(866.0,113.0){\rule[-0.200pt]{0.400pt}{101.660pt}}
\put(220.0,535.0){\rule[-0.200pt]{155.621pt}{0.400pt}}
\put(45,324){\makebox(0,0){$C(i)$}}
\put(543,23){\makebox(0,0){$i$}}
\put(220.0,113.0){\rule[-0.200pt]{0.400pt}{101.660pt}}
\put(220,197){\raisebox{-.8pt}{\makebox(0,0){$\Diamond$}}}
\put(252,195){\raisebox{-.8pt}{\makebox(0,0){$\Diamond$}}}
\put(285,194){\raisebox{-.8pt}{\makebox(0,0){$\Diamond$}}}
\put(317,195){\raisebox{-.8pt}{\makebox(0,0){$\Diamond$}}}
\put(349,196){\raisebox{-.8pt}{\makebox(0,0){$\Diamond$}}}
\put(381,196){\raisebox{-.8pt}{\makebox(0,0){$\Diamond$}}}
\put(414,196){\raisebox{-.8pt}{\makebox(0,0){$\Diamond$}}}
\put(446,196){\raisebox{-.8pt}{\makebox(0,0){$\Diamond$}}}
\put(478,196){\raisebox{-.8pt}{\makebox(0,0){$\Diamond$}}}
\put(511,197){\raisebox{-.8pt}{\makebox(0,0){$\Diamond$}}}
\put(543,198){\raisebox{-.8pt}{\makebox(0,0){$\Diamond$}}}
\put(575,199){\raisebox{-.8pt}{\makebox(0,0){$\Diamond$}}}
\put(608,200){\raisebox{-.8pt}{\makebox(0,0){$\Diamond$}}}
\put(640,201){\raisebox{-.8pt}{\makebox(0,0){$\Diamond$}}}
\put(672,201){\raisebox{-.8pt}{\makebox(0,0){$\Diamond$}}}
\put(704,202){\raisebox{-.8pt}{\makebox(0,0){$\Diamond$}}}
\put(737,204){\raisebox{-.8pt}{\makebox(0,0){$\Diamond$}}}
\put(769,205){\raisebox{-.8pt}{\makebox(0,0){$\Diamond$}}}
\put(801,205){\raisebox{-.8pt}{\makebox(0,0){$\Diamond$}}}
\put(834,206){\raisebox{-.8pt}{\makebox(0,0){$\Diamond$}}}
\put(866,207){\raisebox{-.8pt}{\makebox(0,0){$\Diamond$}}}
\put(220,197){\usebox{\plotpoint}}
\put(220.00,197.00){\usebox{\plotpoint}}
\multiput(227,197)(20.756,0.000){0}{\usebox{\plotpoint}}
\multiput(233,197)(20.756,0.000){0}{\usebox{\plotpoint}}
\put(240.76,197.00){\usebox{\plotpoint}}
\multiput(246,197)(20.756,0.000){0}{\usebox{\plotpoint}}
\multiput(253,197)(20.756,0.000){0}{\usebox{\plotpoint}}
\put(261.51,197.00){\usebox{\plotpoint}}
\multiput(266,197)(20.756,0.000){0}{\usebox{\plotpoint}}
\multiput(272,197)(20.756,0.000){0}{\usebox{\plotpoint}}
\put(282.27,197.00){\usebox{\plotpoint}}
\multiput(285,197)(20.756,0.000){0}{\usebox{\plotpoint}}
\multiput(292,197)(20.756,0.000){0}{\usebox{\plotpoint}}
\put(303.02,197.00){\usebox{\plotpoint}}
\multiput(305,197)(20.756,0.000){0}{\usebox{\plotpoint}}
\multiput(311,197)(20.756,0.000){0}{\usebox{\plotpoint}}
\put(323.78,197.00){\usebox{\plotpoint}}
\multiput(324,197)(20.756,0.000){0}{\usebox{\plotpoint}}
\multiput(331,197)(20.756,0.000){0}{\usebox{\plotpoint}}
\multiput(337,197)(20.756,0.000){0}{\usebox{\plotpoint}}
\put(344.53,197.00){\usebox{\plotpoint}}
\multiput(351,197)(20.756,0.000){0}{\usebox{\plotpoint}}
\multiput(357,197)(20.756,0.000){0}{\usebox{\plotpoint}}
\put(365.29,197.00){\usebox{\plotpoint}}
\multiput(370,197)(20.756,0.000){0}{\usebox{\plotpoint}}
\multiput(377,197)(20.756,0.000){0}{\usebox{\plotpoint}}
\put(386.04,197.00){\usebox{\plotpoint}}
\multiput(390,197)(20.756,0.000){0}{\usebox{\plotpoint}}
\multiput(396,197)(20.756,0.000){0}{\usebox{\plotpoint}}
\put(406.80,197.00){\usebox{\plotpoint}}
\multiput(409,197)(20.756,0.000){0}{\usebox{\plotpoint}}
\multiput(416,197)(20.756,0.000){0}{\usebox{\plotpoint}}
\put(427.55,197.00){\usebox{\plotpoint}}
\multiput(429,197)(20.756,0.000){0}{\usebox{\plotpoint}}
\multiput(435,197)(20.756,0.000){0}{\usebox{\plotpoint}}
\multiput(442,197)(20.756,0.000){0}{\usebox{\plotpoint}}
\put(448.31,197.00){\usebox{\plotpoint}}
\multiput(455,197)(20.756,0.000){0}{\usebox{\plotpoint}}
\multiput(461,197)(20.756,0.000){0}{\usebox{\plotpoint}}
\put(469.07,197.00){\usebox{\plotpoint}}
\multiput(474,197)(20.756,0.000){0}{\usebox{\plotpoint}}
\multiput(481,197)(20.756,0.000){0}{\usebox{\plotpoint}}
\put(489.82,197.00){\usebox{\plotpoint}}
\multiput(494,197)(20.756,0.000){0}{\usebox{\plotpoint}}
\multiput(501,197)(20.756,0.000){0}{\usebox{\plotpoint}}
\put(510.58,197.00){\usebox{\plotpoint}}
\multiput(514,197)(20.756,0.000){0}{\usebox{\plotpoint}}
\multiput(520,197)(20.756,0.000){0}{\usebox{\plotpoint}}
\put(531.33,197.00){\usebox{\plotpoint}}
\multiput(533,197)(20.756,0.000){0}{\usebox{\plotpoint}}
\multiput(540,197)(20.756,0.000){0}{\usebox{\plotpoint}}
\put(552.09,197.00){\usebox{\plotpoint}}
\multiput(553,197)(20.756,0.000){0}{\usebox{\plotpoint}}
\multiput(559,197)(20.756,0.000){0}{\usebox{\plotpoint}}
\multiput(566,197)(20.756,0.000){0}{\usebox{\plotpoint}}
\put(572.84,197.00){\usebox{\plotpoint}}
\multiput(579,197)(20.756,0.000){0}{\usebox{\plotpoint}}
\multiput(585,197)(20.756,0.000){0}{\usebox{\plotpoint}}
\put(593.60,197.00){\usebox{\plotpoint}}
\multiput(598,197)(20.756,0.000){0}{\usebox{\plotpoint}}
\multiput(605,197)(20.756,0.000){0}{\usebox{\plotpoint}}
\put(614.35,197.00){\usebox{\plotpoint}}
\multiput(618,197)(20.756,0.000){0}{\usebox{\plotpoint}}
\multiput(625,197)(20.756,0.000){0}{\usebox{\plotpoint}}
\put(635.11,197.00){\usebox{\plotpoint}}
\multiput(638,197)(20.756,0.000){0}{\usebox{\plotpoint}}
\multiput(644,197)(20.756,0.000){0}{\usebox{\plotpoint}}
\put(655.87,197.00){\usebox{\plotpoint}}
\multiput(657,197)(20.756,0.000){0}{\usebox{\plotpoint}}
\multiput(664,197)(20.756,0.000){0}{\usebox{\plotpoint}}
\put(676.62,197.00){\usebox{\plotpoint}}
\multiput(677,197)(20.756,0.000){0}{\usebox{\plotpoint}}
\multiput(683,197)(20.756,0.000){0}{\usebox{\plotpoint}}
\multiput(690,197)(20.756,0.000){0}{\usebox{\plotpoint}}
\put(697.38,197.00){\usebox{\plotpoint}}
\multiput(703,197)(20.756,0.000){0}{\usebox{\plotpoint}}
\multiput(709,197)(20.756,0.000){0}{\usebox{\plotpoint}}
\put(718.13,197.00){\usebox{\plotpoint}}
\multiput(722,197)(20.756,0.000){0}{\usebox{\plotpoint}}
\multiput(729,197)(20.756,0.000){0}{\usebox{\plotpoint}}
\put(738.89,197.00){\usebox{\plotpoint}}
\multiput(742,197)(20.756,0.000){0}{\usebox{\plotpoint}}
\multiput(749,197)(20.756,0.000){0}{\usebox{\plotpoint}}
\put(759.64,197.00){\usebox{\plotpoint}}
\multiput(762,197)(20.756,0.000){0}{\usebox{\plotpoint}}
\multiput(768,197)(20.756,0.000){0}{\usebox{\plotpoint}}
\put(780.40,197.00){\usebox{\plotpoint}}
\multiput(781,197)(20.756,0.000){0}{\usebox{\plotpoint}}
\multiput(788,197)(20.756,0.000){0}{\usebox{\plotpoint}}
\multiput(794,197)(20.756,0.000){0}{\usebox{\plotpoint}}
\put(801.15,197.00){\usebox{\plotpoint}}
\multiput(807,197)(20.756,0.000){0}{\usebox{\plotpoint}}
\multiput(814,197)(20.756,0.000){0}{\usebox{\plotpoint}}
\put(821.91,197.00){\usebox{\plotpoint}}
\multiput(827,197)(20.756,0.000){0}{\usebox{\plotpoint}}
\multiput(833,197)(20.756,0.000){0}{\usebox{\plotpoint}}
\put(842.66,197.00){\usebox{\plotpoint}}
\multiput(846,197)(20.756,0.000){0}{\usebox{\plotpoint}}
\multiput(853,197)(20.756,0.000){0}{\usebox{\plotpoint}}
\put(863.42,197.00){\usebox{\plotpoint}}
\put(866,197){\usebox{\plotpoint}}
\vspace{-10mm}
\caption{$C(i)$ as a function of $i$
for $N_f=0$, $\beta=5.5088$ and $V=6^4$ 
using Eq. (6).}
\label{fig2}
\end{figure}

\begin{figure}[htb]
%\fpsxsize=6.2cm
\hspace{6mm}
%\def\fpsangle{270}
% GNUPLOT: LaTeX picture
\setlength{\unitlength}{0.240900pt}
\ifx\plotpoint\undefined\newsavebox{\plotpoint}\fi
\font\gnuplot=cmr10 at 10pt
\sbox{\plotpoint}{\rule[-0.200pt]{0.400pt}{0.400pt}}%
\put(220.0,113.0){\rule[-0.200pt]{4.818pt}{0.400pt}}
\put(198,113){\makebox(0,0)[r]{0.06}}
\put(846.0,113.0){\rule[-0.200pt]{4.818pt}{0.400pt}}
\put(220.0,166.0){\rule[-0.200pt]{4.818pt}{0.400pt}}
\put(198,166){\makebox(0,0)[r]{0.065}}
\put(846.0,166.0){\rule[-0.200pt]{4.818pt}{0.400pt}}
\put(220.0,219.0){\rule[-0.200pt]{4.818pt}{0.400pt}}
\put(198,219){\makebox(0,0)[r]{0.07}}
\put(846.0,219.0){\rule[-0.200pt]{4.818pt}{0.400pt}}
\put(220.0,271.0){\rule[-0.200pt]{4.818pt}{0.400pt}}
\put(198,271){\makebox(0,0)[r]{0.075}}
\put(846.0,271.0){\rule[-0.200pt]{4.818pt}{0.400pt}}
\put(220.0,324.0){\rule[-0.200pt]{4.818pt}{0.400pt}}
\put(198,324){\makebox(0,0)[r]{0.08}}
\put(846.0,324.0){\rule[-0.200pt]{4.818pt}{0.400pt}}
\put(220.0,377.0){\rule[-0.200pt]{4.818pt}{0.400pt}}
\put(198,377){\makebox(0,0)[r]{0.085}}
\put(846.0,377.0){\rule[-0.200pt]{4.818pt}{0.400pt}}
\put(220.0,430.0){\rule[-0.200pt]{4.818pt}{0.400pt}}
\put(198,430){\makebox(0,0)[r]{0.09}}
\put(846.0,430.0){\rule[-0.200pt]{4.818pt}{0.400pt}}
\put(220.0,482.0){\rule[-0.200pt]{4.818pt}{0.400pt}}
\put(198,482){\makebox(0,0)[r]{0.095}}
\put(846.0,482.0){\rule[-0.200pt]{4.818pt}{0.400pt}}
\put(220.0,535.0){\rule[-0.200pt]{4.818pt}{0.400pt}}
\put(198,535){\makebox(0,0)[r]{0.1}}
\put(846.0,535.0){\rule[-0.200pt]{4.818pt}{0.400pt}}
\put(220.0,113.0){\rule[-0.200pt]{0.400pt}{4.818pt}}
\put(220,68){\makebox(0,0){10}}
\put(220.0,515.0){\rule[-0.200pt]{0.400pt}{4.818pt}}
\put(328.0,113.0){\rule[-0.200pt]{0.400pt}{4.818pt}}
\put(328,68){\makebox(0,0){15}}
\put(328.0,515.0){\rule[-0.200pt]{0.400pt}{4.818pt}}
\put(435.0,113.0){\rule[-0.200pt]{0.400pt}{4.818pt}}
\put(435,68){\makebox(0,0){20}}
\put(435.0,515.0){\rule[-0.200pt]{0.400pt}{4.818pt}}
\put(543.0,113.0){\rule[-0.200pt]{0.400pt}{4.818pt}}
\put(543,68){\makebox(0,0){25}}
\put(543.0,515.0){\rule[-0.200pt]{0.400pt}{4.818pt}}
\put(651.0,113.0){\rule[-0.200pt]{0.400pt}{4.818pt}}
\put(651,68){\makebox(0,0){30}}
\put(651.0,515.0){\rule[-0.200pt]{0.400pt}{4.818pt}}
\put(758.0,113.0){\rule[-0.200pt]{0.400pt}{4.818pt}}
\put(758,68){\makebox(0,0){35}}
\put(758.0,515.0){\rule[-0.200pt]{0.400pt}{4.818pt}}
\put(866.0,113.0){\rule[-0.200pt]{0.400pt}{4.818pt}}
\put(866,68){\makebox(0,0){40}}
\put(866.0,515.0){\rule[-0.200pt]{0.400pt}{4.818pt}}
\put(220.0,113.0){\rule[-0.200pt]{155.621pt}{0.400pt}}
\put(866.0,113.0){\rule[-0.200pt]{0.400pt}{101.660pt}}
\put(220.0,535.0){\rule[-0.200pt]{155.621pt}{0.400pt}}
\put(45,324){\makebox(0,0){$C(i)$}}
\put(543,23){\makebox(0,0){$i$}}
\put(220.0,113.0){\rule[-0.200pt]{0.400pt}{101.660pt}}
\put(220,273){\raisebox{-.8pt}{\makebox(0,0){$\Diamond$}}}
\put(242,272){\raisebox{-.8pt}{\makebox(0,0){$\Diamond$}}}
\put(263,281){\raisebox{-.8pt}{\makebox(0,0){$\Diamond$}}}
\put(285,279){\raisebox{-.8pt}{\makebox(0,0){$\Diamond$}}}
\put(306,267){\raisebox{-.8pt}{\makebox(0,0){$\Diamond$}}}
\put(328,276){\raisebox{-.8pt}{\makebox(0,0){$\Diamond$}}}
\put(349,269){\raisebox{-.8pt}{\makebox(0,0){$\Diamond$}}}
\put(371,266){\raisebox{-.8pt}{\makebox(0,0){$\Diamond$}}}
\put(392,272){\raisebox{-.8pt}{\makebox(0,0){$\Diamond$}}}
\put(414,270){\raisebox{-.8pt}{\makebox(0,0){$\Diamond$}}}
\put(435,269){\raisebox{-.8pt}{\makebox(0,0){$\Diamond$}}}
\put(457,272){\raisebox{-.8pt}{\makebox(0,0){$\Diamond$}}}
\put(478,270){\raisebox{-.8pt}{\makebox(0,0){$\Diamond$}}}
\put(500,271){\raisebox{-.8pt}{\makebox(0,0){$\Diamond$}}}
\put(521,271){\raisebox{-.8pt}{\makebox(0,0){$\Diamond$}}}
\put(543,271){\raisebox{-.8pt}{\makebox(0,0){$\Diamond$}}}
\put(565,270){\raisebox{-.8pt}{\makebox(0,0){$\Diamond$}}}
\put(586,271){\raisebox{-.8pt}{\makebox(0,0){$\Diamond$}}}
\put(608,270){\raisebox{-.8pt}{\makebox(0,0){$\Diamond$}}}
\put(629,268){\raisebox{-.8pt}{\makebox(0,0){$\Diamond$}}}
\put(651,272){\raisebox{-.8pt}{\makebox(0,0){$\Diamond$}}}
\put(672,269){\raisebox{-.8pt}{\makebox(0,0){$\Diamond$}}}
\put(694,271){\raisebox{-.8pt}{\makebox(0,0){$\Diamond$}}}
\put(715,270){\raisebox{-.8pt}{\makebox(0,0){$\Diamond$}}}
\put(737,272){\raisebox{-.8pt}{\makebox(0,0){$\Diamond$}}}
\put(758,273){\raisebox{-.8pt}{\makebox(0,0){$\Diamond$}}}
\put(780,276){\raisebox{-.8pt}{\makebox(0,0){$\Diamond$}}}
\put(801,275){\raisebox{-.8pt}{\makebox(0,0){$\Diamond$}}}
\put(823,277){\raisebox{-.8pt}{\makebox(0,0){$\Diamond$}}}
\put(844,277){\raisebox{-.8pt}{\makebox(0,0){$\Diamond$}}}
\put(866,278){\raisebox{-.8pt}{\makebox(0,0){$\Diamond$}}}
\put(220.0,260.0){\rule[-0.200pt]{0.400pt}{6.504pt}}
\put(210.0,260.0){\rule[-0.200pt]{4.818pt}{0.400pt}}
\put(210.0,287.0){\rule[-0.200pt]{4.818pt}{0.400pt}}
\put(242.0,262.0){\rule[-0.200pt]{0.400pt}{4.577pt}}
\put(232.0,262.0){\rule[-0.200pt]{4.818pt}{0.400pt}}
\put(232.0,281.0){\rule[-0.200pt]{4.818pt}{0.400pt}}
\put(263.0,273.0){\rule[-0.200pt]{0.400pt}{3.613pt}}
\put(253.0,273.0){\rule[-0.200pt]{4.818pt}{0.400pt}}
\put(253.0,288.0){\rule[-0.200pt]{4.818pt}{0.400pt}}
\put(285.0,274.0){\rule[-0.200pt]{0.400pt}{2.409pt}}
\put(275.0,274.0){\rule[-0.200pt]{4.818pt}{0.400pt}}
\put(275.0,284.0){\rule[-0.200pt]{4.818pt}{0.400pt}}
\put(306.0,260.0){\rule[-0.200pt]{0.400pt}{3.613pt}}
\put(296.0,260.0){\rule[-0.200pt]{4.818pt}{0.400pt}}
\put(296.0,275.0){\rule[-0.200pt]{4.818pt}{0.400pt}}
\put(328.0,271.0){\rule[-0.200pt]{0.400pt}{2.168pt}}
\put(318.0,271.0){\rule[-0.200pt]{4.818pt}{0.400pt}}
\put(318.0,280.0){\rule[-0.200pt]{4.818pt}{0.400pt}}
\put(349.0,263.0){\rule[-0.200pt]{0.400pt}{2.891pt}}
\put(339.0,263.0){\rule[-0.200pt]{4.818pt}{0.400pt}}
\put(339.0,275.0){\rule[-0.200pt]{4.818pt}{0.400pt}}
\put(371.0,262.0){\rule[-0.200pt]{0.400pt}{2.168pt}}
\put(361.0,262.0){\rule[-0.200pt]{4.818pt}{0.400pt}}
\put(361.0,271.0){\rule[-0.200pt]{4.818pt}{0.400pt}}
\put(392.0,269.0){\rule[-0.200pt]{0.400pt}{1.445pt}}
\put(382.0,269.0){\rule[-0.200pt]{4.818pt}{0.400pt}}
\put(382.0,275.0){\rule[-0.200pt]{4.818pt}{0.400pt}}
\put(414.0,266.0){\rule[-0.200pt]{0.400pt}{2.168pt}}
\put(404.0,266.0){\rule[-0.200pt]{4.818pt}{0.400pt}}
\put(404.0,275.0){\rule[-0.200pt]{4.818pt}{0.400pt}}
\put(435.0,263.0){\rule[-0.200pt]{0.400pt}{2.891pt}}
\put(425.0,263.0){\rule[-0.200pt]{4.818pt}{0.400pt}}
\put(425.0,275.0){\rule[-0.200pt]{4.818pt}{0.400pt}}
\put(457.0,268.0){\rule[-0.200pt]{0.400pt}{1.927pt}}
\put(447.0,268.0){\rule[-0.200pt]{4.818pt}{0.400pt}}
\put(447.0,276.0){\rule[-0.200pt]{4.818pt}{0.400pt}}
\put(478.0,265.0){\rule[-0.200pt]{0.400pt}{2.650pt}}
\put(468.0,265.0){\rule[-0.200pt]{4.818pt}{0.400pt}}
\put(468.0,276.0){\rule[-0.200pt]{4.818pt}{0.400pt}}
\put(500.0,268.0){\rule[-0.200pt]{0.400pt}{1.445pt}}
\put(490.0,268.0){\rule[-0.200pt]{4.818pt}{0.400pt}}
\put(490.0,274.0){\rule[-0.200pt]{4.818pt}{0.400pt}}
\put(521.0,267.0){\rule[-0.200pt]{0.400pt}{2.168pt}}
\put(511.0,267.0){\rule[-0.200pt]{4.818pt}{0.400pt}}
\put(511.0,276.0){\rule[-0.200pt]{4.818pt}{0.400pt}}
\put(543.0,267.0){\rule[-0.200pt]{0.400pt}{1.927pt}}
\put(533.0,267.0){\rule[-0.200pt]{4.818pt}{0.400pt}}
\put(533.0,275.0){\rule[-0.200pt]{4.818pt}{0.400pt}}
\put(565.0,264.0){\rule[-0.200pt]{0.400pt}{2.891pt}}
\put(555.0,264.0){\rule[-0.200pt]{4.818pt}{0.400pt}}
\put(555.0,276.0){\rule[-0.200pt]{4.818pt}{0.400pt}}
\put(586.0,267.0){\rule[-0.200pt]{0.400pt}{2.168pt}}
\put(576.0,267.0){\rule[-0.200pt]{4.818pt}{0.400pt}}
\put(576.0,276.0){\rule[-0.200pt]{4.818pt}{0.400pt}}
\put(608.0,265.0){\rule[-0.200pt]{0.400pt}{2.650pt}}
\put(598.0,265.0){\rule[-0.200pt]{4.818pt}{0.400pt}}
\put(598.0,276.0){\rule[-0.200pt]{4.818pt}{0.400pt}}
\put(629.0,261.0){\rule[-0.200pt]{0.400pt}{3.613pt}}
\put(619.0,261.0){\rule[-0.200pt]{4.818pt}{0.400pt}}
\put(619.0,276.0){\rule[-0.200pt]{4.818pt}{0.400pt}}
\put(651.0,267.0){\rule[-0.200pt]{0.400pt}{2.409pt}}
\put(641.0,267.0){\rule[-0.200pt]{4.818pt}{0.400pt}}
\put(641.0,277.0){\rule[-0.200pt]{4.818pt}{0.400pt}}
\put(672.0,262.0){\rule[-0.200pt]{0.400pt}{3.373pt}}
\put(662.0,262.0){\rule[-0.200pt]{4.818pt}{0.400pt}}
\put(662.0,276.0){\rule[-0.200pt]{4.818pt}{0.400pt}}
\put(694.0,265.0){\rule[-0.200pt]{0.400pt}{3.132pt}}
\put(684.0,265.0){\rule[-0.200pt]{4.818pt}{0.400pt}}
\put(684.0,278.0){\rule[-0.200pt]{4.818pt}{0.400pt}}
\put(715.0,264.0){\rule[-0.200pt]{0.400pt}{3.132pt}}
\put(705.0,264.0){\rule[-0.200pt]{4.818pt}{0.400pt}}
\put(705.0,277.0){\rule[-0.200pt]{4.818pt}{0.400pt}}
\put(737.0,265.0){\rule[-0.200pt]{0.400pt}{3.132pt}}
\put(727.0,265.0){\rule[-0.200pt]{4.818pt}{0.400pt}}
\put(727.0,278.0){\rule[-0.200pt]{4.818pt}{0.400pt}}
\put(758.0,267.0){\rule[-0.200pt]{0.400pt}{2.891pt}}
\put(748.0,267.0){\rule[-0.200pt]{4.818pt}{0.400pt}}
\put(748.0,279.0){\rule[-0.200pt]{4.818pt}{0.400pt}}
\put(780.0,271.0){\rule[-0.200pt]{0.400pt}{2.168pt}}
\put(770.0,271.0){\rule[-0.200pt]{4.818pt}{0.400pt}}
\put(770.0,280.0){\rule[-0.200pt]{4.818pt}{0.400pt}}
\put(801.0,269.0){\rule[-0.200pt]{0.400pt}{2.891pt}}
\put(791.0,269.0){\rule[-0.200pt]{4.818pt}{0.400pt}}
\put(791.0,281.0){\rule[-0.200pt]{4.818pt}{0.400pt}}
\put(823.0,272.0){\rule[-0.200pt]{0.400pt}{2.168pt}}
\put(813.0,272.0){\rule[-0.200pt]{4.818pt}{0.400pt}}
\put(813.0,281.0){\rule[-0.200pt]{4.818pt}{0.400pt}}
\put(844.0,271.0){\rule[-0.200pt]{0.400pt}{2.650pt}}
\put(834.0,271.0){\rule[-0.200pt]{4.818pt}{0.400pt}}
\put(834.0,282.0){\rule[-0.200pt]{4.818pt}{0.400pt}}
\put(866.0,273.0){\rule[-0.200pt]{0.400pt}{2.650pt}}
\put(856.0,273.0){\rule[-0.200pt]{4.818pt}{0.400pt}}
\put(856.0,284.0){\rule[-0.200pt]{4.818pt}{0.400pt}}
\put(220,272){\usebox{\plotpoint}}
\put(220.00,272.00){\usebox{\plotpoint}}
\multiput(227,272)(20.756,0.000){0}{\usebox{\plotpoint}}
\multiput(233,272)(20.756,0.000){0}{\usebox{\plotpoint}}
\put(240.76,272.00){\usebox{\plotpoint}}
\multiput(246,272)(20.756,0.000){0}{\usebox{\plotpoint}}
\multiput(253,272)(20.756,0.000){0}{\usebox{\plotpoint}}
\put(261.51,272.00){\usebox{\plotpoint}}
\multiput(266,272)(20.756,0.000){0}{\usebox{\plotpoint}}
\multiput(272,272)(20.756,0.000){0}{\usebox{\plotpoint}}
\put(282.27,272.00){\usebox{\plotpoint}}
\multiput(285,272)(20.756,0.000){0}{\usebox{\plotpoint}}
\multiput(292,272)(20.756,0.000){0}{\usebox{\plotpoint}}
\put(303.02,272.00){\usebox{\plotpoint}}
\multiput(305,272)(20.756,0.000){0}{\usebox{\plotpoint}}
\multiput(311,272)(20.756,0.000){0}{\usebox{\plotpoint}}
\put(323.78,272.00){\usebox{\plotpoint}}
\multiput(324,272)(20.756,0.000){0}{\usebox{\plotpoint}}
\multiput(331,272)(20.756,0.000){0}{\usebox{\plotpoint}}
\multiput(337,272)(20.756,0.000){0}{\usebox{\plotpoint}}
\put(344.53,272.00){\usebox{\plotpoint}}
\multiput(351,272)(20.756,0.000){0}{\usebox{\plotpoint}}
\multiput(357,272)(20.756,0.000){0}{\usebox{\plotpoint}}
\put(365.29,272.00){\usebox{\plotpoint}}
\multiput(370,272)(20.756,0.000){0}{\usebox{\plotpoint}}
\multiput(377,272)(20.756,0.000){0}{\usebox{\plotpoint}}
\put(386.04,272.00){\usebox{\plotpoint}}
\multiput(390,272)(20.756,0.000){0}{\usebox{\plotpoint}}
\multiput(396,272)(20.756,0.000){0}{\usebox{\plotpoint}}
\put(406.80,272.00){\usebox{\plotpoint}}
\multiput(409,272)(20.756,0.000){0}{\usebox{\plotpoint}}
\multiput(416,272)(20.756,0.000){0}{\usebox{\plotpoint}}
\put(427.55,272.00){\usebox{\plotpoint}}
\multiput(429,272)(20.756,0.000){0}{\usebox{\plotpoint}}
\multiput(435,272)(20.756,0.000){0}{\usebox{\plotpoint}}
\multiput(442,272)(20.756,0.000){0}{\usebox{\plotpoint}}
\put(448.31,272.00){\usebox{\plotpoint}}
\multiput(455,272)(20.756,0.000){0}{\usebox{\plotpoint}}
\multiput(461,272)(20.756,0.000){0}{\usebox{\plotpoint}}
\put(469.07,272.00){\usebox{\plotpoint}}
\multiput(474,272)(20.756,0.000){0}{\usebox{\plotpoint}}
\multiput(481,272)(20.756,0.000){0}{\usebox{\plotpoint}}
\put(489.82,272.00){\usebox{\plotpoint}}
\multiput(494,272)(20.756,0.000){0}{\usebox{\plotpoint}}
\multiput(501,272)(20.756,0.000){0}{\usebox{\plotpoint}}
\put(510.58,272.00){\usebox{\plotpoint}}
\multiput(514,272)(20.756,0.000){0}{\usebox{\plotpoint}}
\multiput(520,272)(20.756,0.000){0}{\usebox{\plotpoint}}
\put(531.33,272.00){\usebox{\plotpoint}}
\multiput(533,272)(20.756,0.000){0}{\usebox{\plotpoint}}
\multiput(540,272)(20.756,0.000){0}{\usebox{\plotpoint}}
\put(552.09,272.00){\usebox{\plotpoint}}
\multiput(553,272)(20.756,0.000){0}{\usebox{\plotpoint}}
\multiput(559,272)(20.756,0.000){0}{\usebox{\plotpoint}}
\multiput(566,272)(20.756,0.000){0}{\usebox{\plotpoint}}
\put(572.84,272.00){\usebox{\plotpoint}}
\multiput(579,272)(20.756,0.000){0}{\usebox{\plotpoint}}
\multiput(585,272)(20.756,0.000){0}{\usebox{\plotpoint}}
\put(593.60,272.00){\usebox{\plotpoint}}
\multiput(598,272)(20.756,0.000){0}{\usebox{\plotpoint}}
\multiput(605,272)(20.756,0.000){0}{\usebox{\plotpoint}}
\put(614.35,272.00){\usebox{\plotpoint}}
\multiput(618,272)(20.756,0.000){0}{\usebox{\plotpoint}}
\multiput(625,272)(20.756,0.000){0}{\usebox{\plotpoint}}
\put(635.11,272.00){\usebox{\plotpoint}}
\multiput(638,272)(20.756,0.000){0}{\usebox{\plotpoint}}
\multiput(644,272)(20.756,0.000){0}{\usebox{\plotpoint}}
\put(655.87,272.00){\usebox{\plotpoint}}
\multiput(657,272)(20.756,0.000){0}{\usebox{\plotpoint}}
\multiput(664,272)(20.756,0.000){0}{\usebox{\plotpoint}}
\put(676.62,272.00){\usebox{\plotpoint}}
\multiput(677,272)(20.756,0.000){0}{\usebox{\plotpoint}}
\multiput(683,272)(20.756,0.000){0}{\usebox{\plotpoint}}
\multiput(690,272)(20.756,0.000){0}{\usebox{\plotpoint}}
\put(697.38,272.00){\usebox{\plotpoint}}
\multiput(703,272)(20.756,0.000){0}{\usebox{\plotpoint}}
\multiput(709,272)(20.756,0.000){0}{\usebox{\plotpoint}}
\put(718.13,272.00){\usebox{\plotpoint}}
\multiput(722,272)(20.756,0.000){0}{\usebox{\plotpoint}}
\multiput(729,272)(20.756,0.000){0}{\usebox{\plotpoint}}
\put(738.89,272.00){\usebox{\plotpoint}}
\multiput(742,272)(20.756,0.000){0}{\usebox{\plotpoint}}
\multiput(749,272)(20.756,0.000){0}{\usebox{\plotpoint}}
\put(759.64,272.00){\usebox{\plotpoint}}
\multiput(762,272)(20.756,0.000){0}{\usebox{\plotpoint}}
\multiput(768,272)(20.756,0.000){0}{\usebox{\plotpoint}}
\put(780.40,272.00){\usebox{\plotpoint}}
\multiput(781,272)(20.756,0.000){0}{\usebox{\plotpoint}}
\multiput(788,272)(20.756,0.000){0}{\usebox{\plotpoint}}
\multiput(794,272)(20.756,0.000){0}{\usebox{\plotpoint}}
\put(801.15,272.00){\usebox{\plotpoint}}
\multiput(807,272)(20.756,0.000){0}{\usebox{\plotpoint}}
\multiput(814,272)(20.756,0.000){0}{\usebox{\plotpoint}}
\put(821.91,272.00){\usebox{\plotpoint}}
\multiput(827,272)(20.756,0.000){0}{\usebox{\plotpoint}}
\multiput(833,272)(20.756,0.000){0}{\usebox{\plotpoint}}
\put(842.66,272.00){\usebox{\plotpoint}}
\multiput(846,272)(20.756,0.000){0}{\usebox{\plotpoint}}
\multiput(853,272)(20.756,0.000){0}{\usebox{\plotpoint}}
\put(863.42,272.00){\usebox{\plotpoint}}
\put(866,272){\usebox{\plotpoint}}
\vspace{-10mm}
\caption{$C(i)$ as a function of $i$
for $N_f=1$, $\beta=5.3469$ and $V=6^4$ 
using Eq. (6).}
\label{fig3}
\end{figure}

\begin{figure}[htb]
%\fpsxsize=6.2cm
\hspace{6mm}
%\def\fpsangle{270}
% GNUPLOT: LaTeX picture
\setlength{\unitlength}{0.240900pt}
\ifx\plotpoint\undefined\newsavebox{\plotpoint}\fi
\font\gnuplot=cmr10 at 10pt
\sbox{\plotpoint}{\rule[-0.200pt]{0.400pt}{0.400pt}}%
\put(220.0,113.0){\rule[-0.200pt]{4.818pt}{0.400pt}}
\put(198,113){\makebox(0,0)[r]{0.06}}
\put(846.0,113.0){\rule[-0.200pt]{4.818pt}{0.400pt}}
\put(220.0,166.0){\rule[-0.200pt]{4.818pt}{0.400pt}}
\put(198,166){\makebox(0,0)[r]{0.065}}
\put(846.0,166.0){\rule[-0.200pt]{4.818pt}{0.400pt}}
\put(220.0,219.0){\rule[-0.200pt]{4.818pt}{0.400pt}}
\put(198,219){\makebox(0,0)[r]{0.07}}
\put(846.0,219.0){\rule[-0.200pt]{4.818pt}{0.400pt}}
\put(220.0,271.0){\rule[-0.200pt]{4.818pt}{0.400pt}}
\put(198,271){\makebox(0,0)[r]{0.075}}
\put(846.0,271.0){\rule[-0.200pt]{4.818pt}{0.400pt}}
\put(220.0,324.0){\rule[-0.200pt]{4.818pt}{0.400pt}}
\put(198,324){\makebox(0,0)[r]{0.08}}
\put(846.0,324.0){\rule[-0.200pt]{4.818pt}{0.400pt}}
\put(220.0,377.0){\rule[-0.200pt]{4.818pt}{0.400pt}}
\put(198,377){\makebox(0,0)[r]{0.085}}
\put(846.0,377.0){\rule[-0.200pt]{4.818pt}{0.400pt}}
\put(220.0,430.0){\rule[-0.200pt]{4.818pt}{0.400pt}}
\put(198,430){\makebox(0,0)[r]{0.09}}
\put(846.0,430.0){\rule[-0.200pt]{4.818pt}{0.400pt}}
\put(220.0,482.0){\rule[-0.200pt]{4.818pt}{0.400pt}}
\put(198,482){\makebox(0,0)[r]{0.095}}
\put(846.0,482.0){\rule[-0.200pt]{4.818pt}{0.400pt}}
\put(220.0,535.0){\rule[-0.200pt]{4.818pt}{0.400pt}}
\put(198,535){\makebox(0,0)[r]{0.1}}
\put(846.0,535.0){\rule[-0.200pt]{4.818pt}{0.400pt}}
\put(220.0,113.0){\rule[-0.200pt]{0.400pt}{4.818pt}}
\put(220,68){\makebox(0,0){10}}
\put(220.0,515.0){\rule[-0.200pt]{0.400pt}{4.818pt}}
\put(328.0,113.0){\rule[-0.200pt]{0.400pt}{4.818pt}}
\put(328,68){\makebox(0,0){15}}
\put(328.0,515.0){\rule[-0.200pt]{0.400pt}{4.818pt}}
\put(435.0,113.0){\rule[-0.200pt]{0.400pt}{4.818pt}}
\put(435,68){\makebox(0,0){20}}
\put(435.0,515.0){\rule[-0.200pt]{0.400pt}{4.818pt}}
\put(543.0,113.0){\rule[-0.200pt]{0.400pt}{4.818pt}}
\put(543,68){\makebox(0,0){25}}
\put(543.0,515.0){\rule[-0.200pt]{0.400pt}{4.818pt}}
\put(651.0,113.0){\rule[-0.200pt]{0.400pt}{4.818pt}}
\put(651,68){\makebox(0,0){30}}
\put(651.0,515.0){\rule[-0.200pt]{0.400pt}{4.818pt}}
\put(758.0,113.0){\rule[-0.200pt]{0.400pt}{4.818pt}}
\put(758,68){\makebox(0,0){35}}
\put(758.0,515.0){\rule[-0.200pt]{0.400pt}{4.818pt}}
\put(866.0,113.0){\rule[-0.200pt]{0.400pt}{4.818pt}}
\put(866,68){\makebox(0,0){40}}
\put(866.0,515.0){\rule[-0.200pt]{0.400pt}{4.818pt}}
\put(220.0,113.0){\rule[-0.200pt]{155.621pt}{0.400pt}}
\put(866.0,113.0){\rule[-0.200pt]{0.400pt}{101.660pt}}
\put(220.0,535.0){\rule[-0.200pt]{155.621pt}{0.400pt}}
\put(45,324){\makebox(0,0){$C(i)$}}
\put(543,23){\makebox(0,0){$i$}}
\put(220.0,113.0){\rule[-0.200pt]{0.400pt}{101.660pt}}
\put(220,237){\raisebox{-.8pt}{\makebox(0,0){$\Diamond$}}}
\put(242,243){\raisebox{-.8pt}{\makebox(0,0){$\Diamond$}}}
\put(263,268){\raisebox{-.8pt}{\makebox(0,0){$\Diamond$}}}
\put(285,265){\raisebox{-.8pt}{\makebox(0,0){$\Diamond$}}}
\put(306,246){\raisebox{-.8pt}{\makebox(0,0){$\Diamond$}}}
\put(328,258){\raisebox{-.8pt}{\makebox(0,0){$\Diamond$}}}
\put(349,249){\raisebox{-.8pt}{\makebox(0,0){$\Diamond$}}}
\put(371,247){\raisebox{-.8pt}{\makebox(0,0){$\Diamond$}}}
\put(392,266){\raisebox{-.8pt}{\makebox(0,0){$\Diamond$}}}
\put(414,261){\raisebox{-.8pt}{\makebox(0,0){$\Diamond$}}}
\put(435,247){\raisebox{-.8pt}{\makebox(0,0){$\Diamond$}}}
\put(457,260){\raisebox{-.8pt}{\makebox(0,0){$\Diamond$}}}
\put(478,257){\raisebox{-.8pt}{\makebox(0,0){$\Diamond$}}}
\put(500,257){\raisebox{-.8pt}{\makebox(0,0){$\Diamond$}}}
\put(521,255){\raisebox{-.8pt}{\makebox(0,0){$\Diamond$}}}
\put(543,256){\raisebox{-.8pt}{\makebox(0,0){$\Diamond$}}}
\put(565,247){\raisebox{-.8pt}{\makebox(0,0){$\Diamond$}}}
\put(586,252){\raisebox{-.8pt}{\makebox(0,0){$\Diamond$}}}
\put(608,248){\raisebox{-.8pt}{\makebox(0,0){$\Diamond$}}}
\put(629,242){\raisebox{-.8pt}{\makebox(0,0){$\Diamond$}}}
\put(651,255){\raisebox{-.8pt}{\makebox(0,0){$\Diamond$}}}
\put(672,246){\raisebox{-.8pt}{\makebox(0,0){$\Diamond$}}}
\put(694,248){\raisebox{-.8pt}{\makebox(0,0){$\Diamond$}}}
\put(715,248){\raisebox{-.8pt}{\makebox(0,0){$\Diamond$}}}
\put(737,248){\raisebox{-.8pt}{\makebox(0,0){$\Diamond$}}}
\put(758,250){\raisebox{-.8pt}{\makebox(0,0){$\Diamond$}}}
\put(780,259){\raisebox{-.8pt}{\makebox(0,0){$\Diamond$}}}
\put(801,252){\raisebox{-.8pt}{\makebox(0,0){$\Diamond$}}}
\put(823,258){\raisebox{-.8pt}{\makebox(0,0){$\Diamond$}}}
\put(844,254){\raisebox{-.8pt}{\makebox(0,0){$\Diamond$}}}
\put(866,261){\raisebox{-.8pt}{\makebox(0,0){$\Diamond$}}}
\put(220.0,207.0){\rule[-0.200pt]{0.400pt}{14.213pt}}
\put(210.0,207.0){\rule[-0.200pt]{4.818pt}{0.400pt}}
\put(210.0,266.0){\rule[-0.200pt]{4.818pt}{0.400pt}}
\put(242.0,223.0){\rule[-0.200pt]{0.400pt}{9.636pt}}
\put(232.0,223.0){\rule[-0.200pt]{4.818pt}{0.400pt}}
\put(232.0,263.0){\rule[-0.200pt]{4.818pt}{0.400pt}}
\put(263.0,242.0){\rule[-0.200pt]{0.400pt}{12.527pt}}
\put(253.0,242.0){\rule[-0.200pt]{4.818pt}{0.400pt}}
\put(253.0,294.0){\rule[-0.200pt]{4.818pt}{0.400pt}}
\put(285.0,254.0){\rule[-0.200pt]{0.400pt}{5.541pt}}
\put(275.0,254.0){\rule[-0.200pt]{4.818pt}{0.400pt}}
\put(275.0,277.0){\rule[-0.200pt]{4.818pt}{0.400pt}}
\put(306.0,229.0){\rule[-0.200pt]{0.400pt}{7.950pt}}
\put(296.0,229.0){\rule[-0.200pt]{4.818pt}{0.400pt}}
\put(296.0,262.0){\rule[-0.200pt]{4.818pt}{0.400pt}}
\put(328.0,245.0){\rule[-0.200pt]{0.400pt}{6.504pt}}
\put(318.0,245.0){\rule[-0.200pt]{4.818pt}{0.400pt}}
\put(318.0,272.0){\rule[-0.200pt]{4.818pt}{0.400pt}}
\put(349.0,245.0){\rule[-0.200pt]{0.400pt}{2.168pt}}
\put(339.0,245.0){\rule[-0.200pt]{4.818pt}{0.400pt}}
\put(339.0,254.0){\rule[-0.200pt]{4.818pt}{0.400pt}}
\put(371.0,243.0){\rule[-0.200pt]{0.400pt}{2.168pt}}
\put(361.0,243.0){\rule[-0.200pt]{4.818pt}{0.400pt}}
\put(361.0,252.0){\rule[-0.200pt]{4.818pt}{0.400pt}}
\put(392.0,264.0){\rule[-0.200pt]{0.400pt}{0.723pt}}
\put(382.0,264.0){\rule[-0.200pt]{4.818pt}{0.400pt}}
\put(382.0,267.0){\rule[-0.200pt]{4.818pt}{0.400pt}}
\put(414.0,255.0){\rule[-0.200pt]{0.400pt}{2.650pt}}
\put(404.0,255.0){\rule[-0.200pt]{4.818pt}{0.400pt}}
\put(404.0,266.0){\rule[-0.200pt]{4.818pt}{0.400pt}}
\put(435.0,240.0){\rule[-0.200pt]{0.400pt}{3.373pt}}
\put(425.0,240.0){\rule[-0.200pt]{4.818pt}{0.400pt}}
\put(425.0,254.0){\rule[-0.200pt]{4.818pt}{0.400pt}}
\put(457.0,256.0){\rule[-0.200pt]{0.400pt}{2.168pt}}
\put(447.0,256.0){\rule[-0.200pt]{4.818pt}{0.400pt}}
\put(447.0,265.0){\rule[-0.200pt]{4.818pt}{0.400pt}}
\put(478.0,242.0){\rule[-0.200pt]{0.400pt}{7.468pt}}
\put(468.0,242.0){\rule[-0.200pt]{4.818pt}{0.400pt}}
\put(468.0,273.0){\rule[-0.200pt]{4.818pt}{0.400pt}}
\put(500.0,254.0){\rule[-0.200pt]{0.400pt}{1.445pt}}
\put(490.0,254.0){\rule[-0.200pt]{4.818pt}{0.400pt}}
\put(490.0,260.0){\rule[-0.200pt]{4.818pt}{0.400pt}}
\put(521.0,250.0){\rule[-0.200pt]{0.400pt}{2.409pt}}
\put(511.0,250.0){\rule[-0.200pt]{4.818pt}{0.400pt}}
\put(511.0,260.0){\rule[-0.200pt]{4.818pt}{0.400pt}}
\put(543.0,251.0){\rule[-0.200pt]{0.400pt}{2.409pt}}
\put(533.0,251.0){\rule[-0.200pt]{4.818pt}{0.400pt}}
\put(533.0,261.0){\rule[-0.200pt]{4.818pt}{0.400pt}}
\put(565.0,241.0){\rule[-0.200pt]{0.400pt}{2.891pt}}
\put(555.0,241.0){\rule[-0.200pt]{4.818pt}{0.400pt}}
\put(555.0,253.0){\rule[-0.200pt]{4.818pt}{0.400pt}}
\put(586.0,245.0){\rule[-0.200pt]{0.400pt}{3.373pt}}
\put(576.0,245.0){\rule[-0.200pt]{4.818pt}{0.400pt}}
\put(576.0,259.0){\rule[-0.200pt]{4.818pt}{0.400pt}}
\put(608.0,243.0){\rule[-0.200pt]{0.400pt}{2.409pt}}
\put(598.0,243.0){\rule[-0.200pt]{4.818pt}{0.400pt}}
\put(598.0,253.0){\rule[-0.200pt]{4.818pt}{0.400pt}}
\put(629.0,234.0){\rule[-0.200pt]{0.400pt}{3.854pt}}
\put(619.0,234.0){\rule[-0.200pt]{4.818pt}{0.400pt}}
\put(619.0,250.0){\rule[-0.200pt]{4.818pt}{0.400pt}}
\put(651.0,249.0){\rule[-0.200pt]{0.400pt}{2.891pt}}
\put(641.0,249.0){\rule[-0.200pt]{4.818pt}{0.400pt}}
\put(641.0,261.0){\rule[-0.200pt]{4.818pt}{0.400pt}}
\put(672.0,239.0){\rule[-0.200pt]{0.400pt}{3.132pt}}
\put(662.0,239.0){\rule[-0.200pt]{4.818pt}{0.400pt}}
\put(662.0,252.0){\rule[-0.200pt]{4.818pt}{0.400pt}}
\put(694.0,239.0){\rule[-0.200pt]{0.400pt}{4.336pt}}
\put(684.0,239.0){\rule[-0.200pt]{4.818pt}{0.400pt}}
\put(684.0,257.0){\rule[-0.200pt]{4.818pt}{0.400pt}}
\put(715.0,241.0){\rule[-0.200pt]{0.400pt}{3.132pt}}
\put(705.0,241.0){\rule[-0.200pt]{4.818pt}{0.400pt}}
\put(705.0,254.0){\rule[-0.200pt]{4.818pt}{0.400pt}}
\put(737.0,240.0){\rule[-0.200pt]{0.400pt}{3.613pt}}
\put(727.0,240.0){\rule[-0.200pt]{4.818pt}{0.400pt}}
\put(727.0,255.0){\rule[-0.200pt]{4.818pt}{0.400pt}}
\put(758.0,243.0){\rule[-0.200pt]{0.400pt}{3.613pt}}
\put(748.0,243.0){\rule[-0.200pt]{4.818pt}{0.400pt}}
\put(748.0,258.0){\rule[-0.200pt]{4.818pt}{0.400pt}}
\put(780.0,252.0){\rule[-0.200pt]{0.400pt}{3.373pt}}
\put(770.0,252.0){\rule[-0.200pt]{4.818pt}{0.400pt}}
\put(770.0,266.0){\rule[-0.200pt]{4.818pt}{0.400pt}}
\put(801.0,245.0){\rule[-0.200pt]{0.400pt}{3.132pt}}
\put(791.0,245.0){\rule[-0.200pt]{4.818pt}{0.400pt}}
\put(791.0,258.0){\rule[-0.200pt]{4.818pt}{0.400pt}}
\put(823.0,253.0){\rule[-0.200pt]{0.400pt}{2.168pt}}
\put(813.0,253.0){\rule[-0.200pt]{4.818pt}{0.400pt}}
\put(813.0,262.0){\rule[-0.200pt]{4.818pt}{0.400pt}}
\put(844.0,247.0){\rule[-0.200pt]{0.400pt}{3.373pt}}
\put(834.0,247.0){\rule[-0.200pt]{4.818pt}{0.400pt}}
\put(834.0,261.0){\rule[-0.200pt]{4.818pt}{0.400pt}}
\put(866.0,255.0){\rule[-0.200pt]{0.400pt}{2.891pt}}
\put(856.0,255.0){\rule[-0.200pt]{4.818pt}{0.400pt}}
\put(856.0,267.0){\rule[-0.200pt]{4.818pt}{0.400pt}}
\put(220,253){\usebox{\plotpoint}}
\put(220.00,253.00){\usebox{\plotpoint}}
\multiput(227,253)(20.756,0.000){0}{\usebox{\plotpoint}}
\multiput(233,253)(20.756,0.000){0}{\usebox{\plotpoint}}
\put(240.76,253.00){\usebox{\plotpoint}}
\multiput(246,253)(20.756,0.000){0}{\usebox{\plotpoint}}
\multiput(253,253)(20.756,0.000){0}{\usebox{\plotpoint}}
\put(261.51,253.00){\usebox{\plotpoint}}
\multiput(266,253)(20.756,0.000){0}{\usebox{\plotpoint}}
\multiput(272,253)(20.756,0.000){0}{\usebox{\plotpoint}}
\put(282.27,253.00){\usebox{\plotpoint}}
\multiput(285,253)(20.756,0.000){0}{\usebox{\plotpoint}}
\multiput(292,253)(20.756,0.000){0}{\usebox{\plotpoint}}
\put(303.02,253.00){\usebox{\plotpoint}}
\multiput(305,253)(20.756,0.000){0}{\usebox{\plotpoint}}
\multiput(311,253)(20.756,0.000){0}{\usebox{\plotpoint}}
\put(323.78,253.00){\usebox{\plotpoint}}
\multiput(324,253)(20.756,0.000){0}{\usebox{\plotpoint}}
\multiput(331,253)(20.756,0.000){0}{\usebox{\plotpoint}}
\multiput(337,253)(20.756,0.000){0}{\usebox{\plotpoint}}
\put(344.53,253.00){\usebox{\plotpoint}}
\multiput(351,253)(20.756,0.000){0}{\usebox{\plotpoint}}
\multiput(357,253)(20.756,0.000){0}{\usebox{\plotpoint}}
\put(365.29,253.00){\usebox{\plotpoint}}
\multiput(370,253)(20.756,0.000){0}{\usebox{\plotpoint}}
\multiput(377,253)(20.756,0.000){0}{\usebox{\plotpoint}}
\put(386.04,253.00){\usebox{\plotpoint}}
\multiput(390,253)(20.756,0.000){0}{\usebox{\plotpoint}}
\multiput(396,253)(20.756,0.000){0}{\usebox{\plotpoint}}
\put(406.80,253.00){\usebox{\plotpoint}}
\multiput(409,253)(20.756,0.000){0}{\usebox{\plotpoint}}
\multiput(416,253)(20.756,0.000){0}{\usebox{\plotpoint}}
\put(427.55,253.00){\usebox{\plotpoint}}
\multiput(429,253)(20.756,0.000){0}{\usebox{\plotpoint}}
\multiput(435,253)(20.756,0.000){0}{\usebox{\plotpoint}}
\multiput(442,253)(20.756,0.000){0}{\usebox{\plotpoint}}
\put(448.31,253.00){\usebox{\plotpoint}}
\multiput(455,253)(20.756,0.000){0}{\usebox{\plotpoint}}
\multiput(461,253)(20.756,0.000){0}{\usebox{\plotpoint}}
\put(469.07,253.00){\usebox{\plotpoint}}
\multiput(474,253)(20.756,0.000){0}{\usebox{\plotpoint}}
\multiput(481,253)(20.756,0.000){0}{\usebox{\plotpoint}}
\put(489.82,253.00){\usebox{\plotpoint}}
\multiput(494,253)(20.756,0.000){0}{\usebox{\plotpoint}}
\multiput(501,253)(20.756,0.000){0}{\usebox{\plotpoint}}
\put(510.58,253.00){\usebox{\plotpoint}}
\multiput(514,253)(20.756,0.000){0}{\usebox{\plotpoint}}
\multiput(520,253)(20.756,0.000){0}{\usebox{\plotpoint}}
\put(531.33,253.00){\usebox{\plotpoint}}
\multiput(533,253)(20.756,0.000){0}{\usebox{\plotpoint}}
\multiput(540,253)(20.756,0.000){0}{\usebox{\plotpoint}}
\put(552.09,253.00){\usebox{\plotpoint}}
\multiput(553,253)(20.756,0.000){0}{\usebox{\plotpoint}}
\multiput(559,253)(20.756,0.000){0}{\usebox{\plotpoint}}
\multiput(566,253)(20.756,0.000){0}{\usebox{\plotpoint}}
\put(572.84,253.00){\usebox{\plotpoint}}
\multiput(579,253)(20.756,0.000){0}{\usebox{\plotpoint}}
\multiput(585,253)(20.756,0.000){0}{\usebox{\plotpoint}}
\put(593.60,253.00){\usebox{\plotpoint}}
\multiput(598,253)(20.756,0.000){0}{\usebox{\plotpoint}}
\multiput(605,253)(20.756,0.000){0}{\usebox{\plotpoint}}
\put(614.35,253.00){\usebox{\plotpoint}}
\multiput(618,253)(20.756,0.000){0}{\usebox{\plotpoint}}
\multiput(625,253)(20.756,0.000){0}{\usebox{\plotpoint}}
\put(635.11,253.00){\usebox{\plotpoint}}
\multiput(638,253)(20.756,0.000){0}{\usebox{\plotpoint}}
\multiput(644,253)(20.756,0.000){0}{\usebox{\plotpoint}}
\put(655.87,253.00){\usebox{\plotpoint}}
\multiput(657,253)(20.756,0.000){0}{\usebox{\plotpoint}}
\multiput(664,253)(20.756,0.000){0}{\usebox{\plotpoint}}
\put(676.62,253.00){\usebox{\plotpoint}}
\multiput(677,253)(20.756,0.000){0}{\usebox{\plotpoint}}
\multiput(683,253)(20.756,0.000){0}{\usebox{\plotpoint}}
\multiput(690,253)(20.756,0.000){0}{\usebox{\plotpoint}}
\put(697.38,253.00){\usebox{\plotpoint}}
\multiput(703,253)(20.756,0.000){0}{\usebox{\plotpoint}}
\multiput(709,253)(20.756,0.000){0}{\usebox{\plotpoint}}
\put(718.13,253.00){\usebox{\plotpoint}}
\multiput(722,253)(20.756,0.000){0}{\usebox{\plotpoint}}
\multiput(729,253)(20.756,0.000){0}{\usebox{\plotpoint}}
\put(738.89,253.00){\usebox{\plotpoint}}
\multiput(742,253)(20.756,0.000){0}{\usebox{\plotpoint}}
\multiput(749,253)(20.756,0.000){0}{\usebox{\plotpoint}}
\put(759.64,253.00){\usebox{\plotpoint}}
\multiput(762,253)(20.756,0.000){0}{\usebox{\plotpoint}}
\multiput(768,253)(20.756,0.000){0}{\usebox{\plotpoint}}
\put(780.40,253.00){\usebox{\plotpoint}}
\multiput(781,253)(20.756,0.000){0}{\usebox{\plotpoint}}
\multiput(788,253)(20.756,0.000){0}{\usebox{\plotpoint}}
\multiput(794,253)(20.756,0.000){0}{\usebox{\plotpoint}}
\put(801.15,253.00){\usebox{\plotpoint}}
\multiput(807,253)(20.756,0.000){0}{\usebox{\plotpoint}}
\multiput(814,253)(20.756,0.000){0}{\usebox{\plotpoint}}
\put(821.91,253.00){\usebox{\plotpoint}}
\multiput(827,253)(20.756,0.000){0}{\usebox{\plotpoint}}
\multiput(833,253)(20.756,0.000){0}{\usebox{\plotpoint}}
\put(842.66,253.00){\usebox{\plotpoint}}
\multiput(846,253)(20.756,0.000){0}{\usebox{\plotpoint}}
\multiput(853,253)(20.756,0.000){0}{\usebox{\plotpoint}}
\put(863.42,253.00){\usebox{\plotpoint}}
\put(866,253){\usebox{\plotpoint}}
\vspace{-10mm}
\caption{$C(i)$ as a function of $i$
for $N_f=2$, $\beta=5.3136$ and $V=6^4$ 
using Eq. (6).}
\label{fig4}
\end{figure}

\begin{figure}[htb]
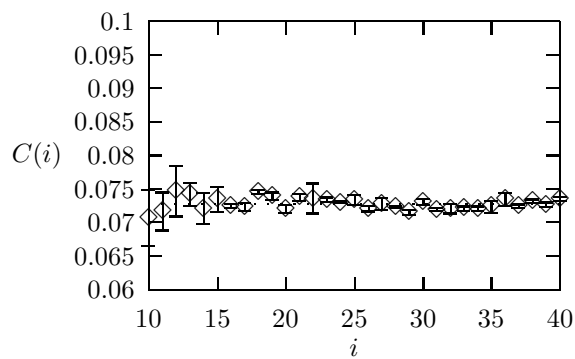

%\fpsxsize=6.2cm
\hspace{6mm}
%\def\fpsangle{270}
% GNUPLOT: LaTeX picture
\setlength{\unitlength}{0.240900pt}
\ifx\plotpoint\undefined\newsavebox{\plotpoint}\fi
\font\gnuplot=cmr10 at 10pt
\sbox{\plotpoint}{\rule[-0.200pt]{0.400pt}{0.400pt}}%
\put(220.0,113.0){\rule[-0.200pt]{4.818pt}{0.400pt}}
\put(198,113){\makebox(0,0)[r]{0.06}}
\put(846.0,113.0){\rule[-0.200pt]{4.818pt}{0.400pt}}
\put(220.0,166.0){\rule[-0.200pt]{4.818pt}{0.400pt}}
\put(198,166){\makebox(0,0)[r]{0.065}}
\put(846.0,166.0){\rule[-0.200pt]{4.818pt}{0.400pt}}
\put(220.0,219.0){\rule[-0.200pt]{4.818pt}{0.400pt}}
\put(198,219){\makebox(0,0)[r]{0.07}}
\put(846.0,219.0){\rule[-0.200pt]{4.818pt}{0.400pt}}
\put(220.0,271.0){\rule[-0.200pt]{4.818pt}{0.400pt}}
\put(198,271){\makebox(0,0)[r]{0.075}}
\put(846.0,271.0){\rule[-0.200pt]{4.818pt}{0.400pt}}
\put(220.0,324.0){\rule[-0.200pt]{4.818pt}{0.400pt}}
\put(198,324){\makebox(0,0)[r]{0.08}}
\put(846.0,324.0){\rule[-0.200pt]{4.818pt}{0.400pt}}
\put(220.0,377.0){\rule[-0.200pt]{4.818pt}{0.400pt}}
\put(198,377){\makebox(0,0)[r]{0.085}}
\put(846.0,377.0){\rule[-0.200pt]{4.818pt}{0.400pt}}
\put(220.0,430.0){\rule[-0.200pt]{4.818pt}{0.400pt}}
\put(198,430){\makebox(0,0)[r]{0.09}}
\put(846.0,430.0){\rule[-0.200pt]{4.818pt}{0.400pt}}
\put(220.0,482.0){\rule[-0.200pt]{4.818pt}{0.400pt}}
\put(198,482){\makebox(0,0)[r]{0.095}}
\put(846.0,482.0){\rule[-0.200pt]{4.818pt}{0.400pt}}
\put(220.0,535.0){\rule[-0.200pt]{4.818pt}{0.400pt}}
\put(198,535){\makebox(0,0)[r]{0.1}}
\put(846.0,535.0){\rule[-0.200pt]{4.818pt}{0.400pt}}
\put(220.0,113.0){\rule[-0.200pt]{0.400pt}{4.818pt}}
\put(220,68){\makebox(0,0){10}}
\put(220.0,515.0){\rule[-0.200pt]{0.400pt}{4.818pt}}
\put(328.0,113.0){\rule[-0.200pt]{0.400pt}{4.818pt}}
\put(328,68){\makebox(0,0){15}}
\put(328.0,515.0){\rule[-0.200pt]{0.400pt}{4.818pt}}
\put(435.0,113.0){\rule[-0.200pt]{0.400pt}{4.818pt}}
\put(435,68){\makebox(0,0){20}}
\put(435.0,515.0){\rule[-0.200pt]{0.400pt}{4.818pt}}
\put(543.0,113.0){\rule[-0.200pt]{0.400pt}{4.818pt}}
\put(543,68){\makebox(0,0){25}}
\put(543.0,515.0){\rule[-0.200pt]{0.400pt}{4.818pt}}
\put(651.0,113.0){\rule[-0.200pt]{0.400pt}{4.818pt}}
\put(651,68){\makebox(0,0){30}}
\put(651.0,515.0){\rule[-0.200pt]{0.400pt}{4.818pt}}
\put(758.0,113.0){\rule[-0.200pt]{0.400pt}{4.818pt}}
\put(758,68){\makebox(0,0){35}}
\put(758.0,515.0){\rule[-0.200pt]{0.400pt}{4.818pt}}
\put(866.0,113.0){\rule[-0.200pt]{0.400pt}{4.818pt}}
\put(866,68){\makebox(0,0){40}}
\put(866.0,515.0){\rule[-0.200pt]{0.400pt}{4.818pt}}
\put(220.0,113.0){\rule[-0.200pt]{155.621pt}{0.400pt}}
\put(866.0,113.0){\rule[-0.200pt]{0.400pt}{101.660pt}}
\put(220.0,535.0){\rule[-0.200pt]{155.621pt}{0.400pt}}
\put(45,324){\makebox(0,0){$C(i)$}}
\put(543,23){\makebox(0,0){$i$}}
\put(220.0,113.0){\rule[-0.200pt]{0.400pt}{101.660pt}}
\put(220,226){\raisebox{-.8pt}{\makebox(0,0){$\Diamond$}}}
\put(242,236){\raisebox{-.8pt}{\makebox(0,0){$\Diamond$}}}
\put(263,268){\raisebox{-.8pt}{\makebox(0,0){$\Diamond$}}}
\put(285,263){\raisebox{-.8pt}{\makebox(0,0){$\Diamond$}}}
\put(306,240){\raisebox{-.8pt}{\makebox(0,0){$\Diamond$}}}
\put(328,255){\raisebox{-.8pt}{\makebox(0,0){$\Diamond$}}}
\put(349,245){\raisebox{-.8pt}{\makebox(0,0){$\Diamond$}}}
\put(371,243){\raisebox{-.8pt}{\makebox(0,0){$\Diamond$}}}
\put(392,266){\raisebox{-.8pt}{\makebox(0,0){$\Diamond$}}}
\put(414,260){\raisebox{-.8pt}{\makebox(0,0){$\Diamond$}}}
\put(435,240){\raisebox{-.8pt}{\makebox(0,0){$\Diamond$}}}
\put(457,258){\raisebox{-.8pt}{\makebox(0,0){$\Diamond$}}}
\put(478,256){\raisebox{-.8pt}{\makebox(0,0){$\Diamond$}}}
\put(500,254){\raisebox{-.8pt}{\makebox(0,0){$\Diamond$}}}
\put(521,250){\raisebox{-.8pt}{\makebox(0,0){$\Diamond$}}}
\put(543,254){\raisebox{-.8pt}{\makebox(0,0){$\Diamond$}}}
\put(565,240){\raisebox{-.8pt}{\makebox(0,0){$\Diamond$}}}
\put(586,248){\raisebox{-.8pt}{\makebox(0,0){$\Diamond$}}}
\put(608,243){\raisebox{-.8pt}{\makebox(0,0){$\Diamond$}}}
\put(629,235){\raisebox{-.8pt}{\makebox(0,0){$\Diamond$}}}
\put(651,251){\raisebox{-.8pt}{\makebox(0,0){$\Diamond$}}}
\put(672,239){\raisebox{-.8pt}{\makebox(0,0){$\Diamond$}}}
\put(694,240){\raisebox{-.8pt}{\makebox(0,0){$\Diamond$}}}
\put(715,241){\raisebox{-.8pt}{\makebox(0,0){$\Diamond$}}}
\put(737,240){\raisebox{-.8pt}{\makebox(0,0){$\Diamond$}}}
\put(758,244){\raisebox{-.8pt}{\makebox(0,0){$\Diamond$}}}
\put(780,255){\raisebox{-.8pt}{\makebox(0,0){$\Diamond$}}}
\put(801,245){\raisebox{-.8pt}{\makebox(0,0){$\Diamond$}}}
\put(823,252){\raisebox{-.8pt}{\makebox(0,0){$\Diamond$}}}
\put(844,246){\raisebox{-.8pt}{\makebox(0,0){$\Diamond$}}}
\put(866,256){\raisebox{-.8pt}{\makebox(0,0){$\Diamond$}}}
\put(220.0,182.0){\rule[-0.200pt]{0.400pt}{21.199pt}}
\put(210.0,182.0){\rule[-0.200pt]{4.818pt}{0.400pt}}
\put(210.0,270.0){\rule[-0.200pt]{4.818pt}{0.400pt}}
\put(242.0,206.0){\rule[-0.200pt]{0.400pt}{14.454pt}}
\put(232.0,206.0){\rule[-0.200pt]{4.818pt}{0.400pt}}
\put(232.0,266.0){\rule[-0.200pt]{4.818pt}{0.400pt}}
\put(263.0,228.0){\rule[-0.200pt]{0.400pt}{19.272pt}}
\put(253.0,228.0){\rule[-0.200pt]{4.818pt}{0.400pt}}
\put(253.0,308.0){\rule[-0.200pt]{4.818pt}{0.400pt}}
\put(285.0,245.0){\rule[-0.200pt]{0.400pt}{8.672pt}}
\put(275.0,245.0){\rule[-0.200pt]{4.818pt}{0.400pt}}
\put(275.0,281.0){\rule[-0.200pt]{4.818pt}{0.400pt}}
\put(306.0,216.0){\rule[-0.200pt]{0.400pt}{11.804pt}}
\put(296.0,216.0){\rule[-0.200pt]{4.818pt}{0.400pt}}
\put(296.0,265.0){\rule[-0.200pt]{4.818pt}{0.400pt}}
\put(328.0,235.0){\rule[-0.200pt]{0.400pt}{9.636pt}}
\put(318.0,235.0){\rule[-0.200pt]{4.818pt}{0.400pt}}
\put(318.0,275.0){\rule[-0.200pt]{4.818pt}{0.400pt}}
\put(349.0,242.0){\rule[-0.200pt]{0.400pt}{1.445pt}}
\put(339.0,242.0){\rule[-0.200pt]{4.818pt}{0.400pt}}
\put(339.0,248.0){\rule[-0.200pt]{4.818pt}{0.400pt}}
\put(371.0,237.0){\rule[-0.200pt]{0.400pt}{2.891pt}}
\put(361.0,237.0){\rule[-0.200pt]{4.818pt}{0.400pt}}
\put(361.0,249.0){\rule[-0.200pt]{4.818pt}{0.400pt}}
\put(392.0,263.0){\rule[-0.200pt]{0.400pt}{1.686pt}}
\put(382.0,263.0){\rule[-0.200pt]{4.818pt}{0.400pt}}
\put(382.0,270.0){\rule[-0.200pt]{4.818pt}{0.400pt}}
\put(414.0,254.0){\rule[-0.200pt]{0.400pt}{2.891pt}}
\put(404.0,254.0){\rule[-0.200pt]{4.818pt}{0.400pt}}
\put(404.0,266.0){\rule[-0.200pt]{4.818pt}{0.400pt}}
\put(435.0,234.0){\rule[-0.200pt]{0.400pt}{2.891pt}}
\put(425.0,234.0){\rule[-0.200pt]{4.818pt}{0.400pt}}
\put(425.0,246.0){\rule[-0.200pt]{4.818pt}{0.400pt}}
\put(457.0,253.0){\rule[-0.200pt]{0.400pt}{2.650pt}}
\put(447.0,253.0){\rule[-0.200pt]{4.818pt}{0.400pt}}
\put(447.0,264.0){\rule[-0.200pt]{4.818pt}{0.400pt}}
\put(478.0,233.0){\rule[-0.200pt]{0.400pt}{11.322pt}}
\put(468.0,233.0){\rule[-0.200pt]{4.818pt}{0.400pt}}
\put(468.0,280.0){\rule[-0.200pt]{4.818pt}{0.400pt}}
\put(500.0,251.0){\rule[-0.200pt]{0.400pt}{1.686pt}}
\put(490.0,251.0){\rule[-0.200pt]{4.818pt}{0.400pt}}
\put(490.0,258.0){\rule[-0.200pt]{4.818pt}{0.400pt}}
\put(521.0,249.0){\rule[-0.200pt]{0.400pt}{0.723pt}}
\put(511.0,249.0){\rule[-0.200pt]{4.818pt}{0.400pt}}
\put(511.0,252.0){\rule[-0.200pt]{4.818pt}{0.400pt}}
\put(543.0,246.0){\rule[-0.200pt]{0.400pt}{3.854pt}}
\put(533.0,246.0){\rule[-0.200pt]{4.818pt}{0.400pt}}
\put(533.0,262.0){\rule[-0.200pt]{4.818pt}{0.400pt}}
\put(565.0,236.0){\rule[-0.200pt]{0.400pt}{2.168pt}}
\put(555.0,236.0){\rule[-0.200pt]{4.818pt}{0.400pt}}
\put(555.0,245.0){\rule[-0.200pt]{4.818pt}{0.400pt}}
\put(586.0,238.0){\rule[-0.200pt]{0.400pt}{4.577pt}}
\put(576.0,238.0){\rule[-0.200pt]{4.818pt}{0.400pt}}
\put(576.0,257.0){\rule[-0.200pt]{4.818pt}{0.400pt}}
\put(608.0,241.0){\rule[-0.200pt]{0.400pt}{0.723pt}}
\put(598.0,241.0){\rule[-0.200pt]{4.818pt}{0.400pt}}
\put(598.0,244.0){\rule[-0.200pt]{4.818pt}{0.400pt}}
\put(629.0,231.0){\rule[-0.200pt]{0.400pt}{1.686pt}}
\put(619.0,231.0){\rule[-0.200pt]{4.818pt}{0.400pt}}
\put(619.0,238.0){\rule[-0.200pt]{4.818pt}{0.400pt}}
\put(651.0,246.0){\rule[-0.200pt]{0.400pt}{2.409pt}}
\put(641.0,246.0){\rule[-0.200pt]{4.818pt}{0.400pt}}
\put(641.0,256.0){\rule[-0.200pt]{4.818pt}{0.400pt}}
\put(672.0,237.0){\rule[-0.200pt]{0.400pt}{0.964pt}}
\put(662.0,237.0){\rule[-0.200pt]{4.818pt}{0.400pt}}
\put(662.0,241.0){\rule[-0.200pt]{4.818pt}{0.400pt}}
\put(694.0,233.0){\rule[-0.200pt]{0.400pt}{3.613pt}}
\put(684.0,233.0){\rule[-0.200pt]{4.818pt}{0.400pt}}
\put(684.0,248.0){\rule[-0.200pt]{4.818pt}{0.400pt}}
\put(715.0,237.0){\rule[-0.200pt]{0.400pt}{1.686pt}}
\put(705.0,237.0){\rule[-0.200pt]{4.818pt}{0.400pt}}
\put(705.0,244.0){\rule[-0.200pt]{4.818pt}{0.400pt}}
\put(737.0,237.0){\rule[-0.200pt]{0.400pt}{1.686pt}}
\put(727.0,237.0){\rule[-0.200pt]{4.818pt}{0.400pt}}
\put(727.0,244.0){\rule[-0.200pt]{4.818pt}{0.400pt}}
\put(758.0,234.0){\rule[-0.200pt]{0.400pt}{4.577pt}}
\put(748.0,234.0){\rule[-0.200pt]{4.818pt}{0.400pt}}
\put(748.0,253.0){\rule[-0.200pt]{4.818pt}{0.400pt}}
\put(780.0,245.0){\rule[-0.200pt]{0.400pt}{4.818pt}}
\put(770.0,245.0){\rule[-0.200pt]{4.818pt}{0.400pt}}
\put(770.0,265.0){\rule[-0.200pt]{4.818pt}{0.400pt}}
\put(801.0,242.0){\rule[-0.200pt]{0.400pt}{1.204pt}}
\put(791.0,242.0){\rule[-0.200pt]{4.818pt}{0.400pt}}
\put(791.0,247.0){\rule[-0.200pt]{4.818pt}{0.400pt}}
\put(823.0,249.0){\rule[-0.200pt]{0.400pt}{1.445pt}}
\put(813.0,249.0){\rule[-0.200pt]{4.818pt}{0.400pt}}
\put(813.0,255.0){\rule[-0.200pt]{4.818pt}{0.400pt}}
\put(844.0,243.0){\rule[-0.200pt]{0.400pt}{1.686pt}}
\put(834.0,243.0){\rule[-0.200pt]{4.818pt}{0.400pt}}
\put(834.0,250.0){\rule[-0.200pt]{4.818pt}{0.400pt}}
\put(866.0,253.0){\rule[-0.200pt]{0.400pt}{1.445pt}}
\put(856.0,253.0){\rule[-0.200pt]{4.818pt}{0.400pt}}
\put(856.0,259.0){\rule[-0.200pt]{4.818pt}{0.400pt}}
\put(220,248){\usebox{\plotpoint}}
\put(220.00,248.00){\usebox{\plotpoint}}
\multiput(227,248)(20.756,0.000){0}{\usebox{\plotpoint}}
\multiput(233,248)(20.756,0.000){0}{\usebox{\plotpoint}}
\put(240.76,248.00){\usebox{\plotpoint}}
\multiput(246,248)(20.756,0.000){0}{\usebox{\plotpoint}}
\multiput(253,248)(20.756,0.000){0}{\usebox{\plotpoint}}
\put(261.51,248.00){\usebox{\plotpoint}}
\multiput(266,248)(20.756,0.000){0}{\usebox{\plotpoint}}
\multiput(272,248)(20.756,0.000){0}{\usebox{\plotpoint}}
\put(282.27,248.00){\usebox{\plotpoint}}
\multiput(285,248)(20.756,0.000){0}{\usebox{\plotpoint}}
\multiput(292,248)(20.756,0.000){0}{\usebox{\plotpoint}}
\put(303.02,248.00){\usebox{\plotpoint}}
\multiput(305,248)(20.756,0.000){0}{\usebox{\plotpoint}}
\multiput(311,248)(20.756,0.000){0}{\usebox{\plotpoint}}
\put(323.78,248.00){\usebox{\plotpoint}}
\multiput(324,248)(20.756,0.000){0}{\usebox{\plotpoint}}
\multiput(331,248)(20.756,0.000){0}{\usebox{\plotpoint}}
\multiput(337,248)(20.756,0.000){0}{\usebox{\plotpoint}}
\put(344.53,248.00){\usebox{\plotpoint}}
\multiput(351,248)(20.756,0.000){0}{\usebox{\plotpoint}}
\multiput(357,248)(20.756,0.000){0}{\usebox{\plotpoint}}
\put(365.29,248.00){\usebox{\plotpoint}}
\multiput(370,248)(20.756,0.000){0}{\usebox{\plotpoint}}
\multiput(377,248)(20.756,0.000){0}{\usebox{\plotpoint}}
\put(386.04,248.00){\usebox{\plotpoint}}
\multiput(390,248)(20.756,0.000){0}{\usebox{\plotpoint}}
\multiput(396,248)(20.756,0.000){0}{\usebox{\plotpoint}}
\put(406.80,248.00){\usebox{\plotpoint}}
\multiput(409,248)(20.756,0.000){0}{\usebox{\plotpoint}}
\multiput(416,248)(20.756,0.000){0}{\usebox{\plotpoint}}
\put(427.55,248.00){\usebox{\plotpoint}}
\multiput(429,248)(20.756,0.000){0}{\usebox{\plotpoint}}
\multiput(435,248)(20.756,0.000){0}{\usebox{\plotpoint}}
\multiput(442,248)(20.756,0.000){0}{\usebox{\plotpoint}}
\put(448.31,248.00){\usebox{\plotpoint}}
\multiput(455,248)(20.756,0.000){0}{\usebox{\plotpoint}}
\multiput(461,248)(20.756,0.000){0}{\usebox{\plotpoint}}
\put(469.07,248.00){\usebox{\plotpoint}}
\multiput(474,248)(20.756,0.000){0}{\usebox{\plotpoint}}
\multiput(481,248)(20.756,0.000){0}{\usebox{\plotpoint}}
\put(489.82,248.00){\usebox{\plotpoint}}
\multiput(494,248)(20.756,0.000){0}{\usebox{\plotpoint}}
\multiput(501,248)(20.756,0.000){0}{\usebox{\plotpoint}}
\put(510.58,248.00){\usebox{\plotpoint}}
\multiput(514,248)(20.756,0.000){0}{\usebox{\plotpoint}}
\multiput(520,248)(20.756,0.000){0}{\usebox{\plotpoint}}
\put(531.33,248.00){\usebox{\plotpoint}}
\multiput(533,248)(20.756,0.000){0}{\usebox{\plotpoint}}
\multiput(540,248)(20.756,0.000){0}{\usebox{\plotpoint}}
\put(552.09,248.00){\usebox{\plotpoint}}
\multiput(553,248)(20.756,0.000){0}{\usebox{\plotpoint}}
\multiput(559,248)(20.756,0.000){0}{\usebox{\plotpoint}}
\multiput(566,248)(20.756,0.000){0}{\usebox{\plotpoint}}
\put(572.84,248.00){\usebox{\plotpoint}}
\multiput(579,248)(20.756,0.000){0}{\usebox{\plotpoint}}
\multiput(585,248)(20.756,0.000){0}{\usebox{\plotpoint}}
\put(593.60,248.00){\usebox{\plotpoint}}
\multiput(598,248)(20.756,0.000){0}{\usebox{\plotpoint}}
\multiput(605,248)(20.756,0.000){0}{\usebox{\plotpoint}}
\put(614.35,248.00){\usebox{\plotpoint}}
\multiput(618,248)(20.756,0.000){0}{\usebox{\plotpoint}}
\multiput(625,248)(20.756,0.000){0}{\usebox{\plotpoint}}
\put(635.11,248.00){\usebox{\plotpoint}}
\multiput(638,248)(20.756,0.000){0}{\usebox{\plotpoint}}
\multiput(644,248)(20.756,0.000){0}{\usebox{\plotpoint}}
\put(655.87,248.00){\usebox{\plotpoint}}
\multiput(657,248)(20.756,0.000){0}{\usebox{\plotpoint}}
\multiput(664,248)(20.756,0.000){0}{\usebox{\plotpoint}}
\put(676.62,248.00){\usebox{\plotpoint}}
\multiput(677,248)(20.756,0.000){0}{\usebox{\plotpoint}}
\multiput(683,248)(20.756,0.000){0}{\usebox{\plotpoint}}
\multiput(690,248)(20.756,0.000){0}{\usebox{\plotpoint}}
\put(697.38,248.00){\usebox{\plotpoint}}
\multiput(703,248)(20.756,0.000){0}{\usebox{\plotpoint}}
\multiput(709,248)(20.756,0.000){0}{\usebox{\plotpoint}}
\put(718.13,248.00){\usebox{\plotpoint}}
\multiput(722,248)(20.756,0.000){0}{\usebox{\plotpoint}}
\multiput(729,248)(20.756,0.000){0}{\usebox{\plotpoint}}
\put(738.89,248.00){\usebox{\plotpoint}}
\multiput(742,248)(20.756,0.000){0}{\usebox{\plotpoint}}
\multiput(749,248)(20.756,0.000){0}{\usebox{\plotpoint}}
\put(759.64,248.00){\usebox{\plotpoint}}
\multiput(762,248)(20.756,0.000){0}{\usebox{\plotpoint}}
\multiput(768,248)(20.756,0.000){0}{\usebox{\plotpoint}}
\put(780.40,248.00){\usebox{\plotpoint}}
\multiput(781,248)(20.756,0.000){0}{\usebox{\plotpoint}}
\multiput(788,248)(20.756,0.000){0}{\usebox{\plotpoint}}
\multiput(794,248)(20.756,0.000){0}{\usebox{\plotpoint}}
\put(801.15,248.00){\usebox{\plotpoint}}
\multiput(807,248)(20.756,0.000){0}{\usebox{\plotpoint}}
\multiput(814,248)(20.756,0.000){0}{\usebox{\plotpoint}}
\put(821.91,248.00){\usebox{\plotpoint}}
\multiput(827,248)(20.756,0.000){0}{\usebox{\plotpoint}}
\multiput(833,248)(20.756,0.000){0}{\usebox{\plotpoint}}
\put(842.66,248.00){\usebox{\plotpoint}}
\multiput(846,248)(20.756,0.000){0}{\usebox{\plotpoint}}
\multiput(853,248)(20.756,0.000){0}{\usebox{\plotpoint}}
\put(863.42,248.00){\usebox{\plotpoint}}
\put(866,248){\usebox{\plotpoint}}
\vspace{-10mm}
\caption{$C(i)$ as a function of $i$
for $N_f=3$, $\beta=5.2814$ and $V=6^4$ 
using Eq. (6).}
\label{fig5}
\end{figure}

\begin{figure}[htb]
%\fpsxsize=6.2cm
\hspace{6mm}
%\def\fpsangle{270}
%insert the figure here
% GNUPLOT: LaTeX picture
\setlength{\unitlength}{0.240900pt}
\ifx\plotpoint\undefined\newsavebox{\plotpoint}\fi
%\begin{picture}(930,558)(0,0)
\font\gnuplot=cmr10 at 10pt
%\gnuplot
\sbox{\plotpoint}{\rule[-0.200pt]{0.400pt}{0.400pt}}%
\put(220.0,113.0){\rule[-0.200pt]{4.818pt}{0.400pt}}
\put(198,113){\makebox(0,0)[r]{0.06}}
\put(846.0,113.0){\rule[-0.200pt]{4.818pt}{0.400pt}}
\put(220.0,166.0){\rule[-0.200pt]{4.818pt}{0.400pt}}
\put(198,166){\makebox(0,0)[r]{0.065}}
\put(846.0,166.0){\rule[-0.200pt]{4.818pt}{0.400pt}}
\put(220.0,219.0){\rule[-0.200pt]{4.818pt}{0.400pt}}
\put(198,219){\makebox(0,0)[r]{0.07}}
\put(846.0,219.0){\rule[-0.200pt]{4.818pt}{0.400pt}}
\put(220.0,271.0){\rule[-0.200pt]{4.818pt}{0.400pt}}
\put(198,271){\makebox(0,0)[r]{0.075}}
\put(846.0,271.0){\rule[-0.200pt]{4.818pt}{0.400pt}}
\put(220.0,324.0){\rule[-0.200pt]{4.818pt}{0.400pt}}
\put(198,324){\makebox(0,0)[r]{0.08}}
\put(846.0,324.0){\rule[-0.200pt]{4.818pt}{0.400pt}}
\put(220.0,377.0){\rule[-0.200pt]{4.818pt}{0.400pt}}
\put(198,377){\makebox(0,0)[r]{0.085}}
\put(846.0,377.0){\rule[-0.200pt]{4.818pt}{0.400pt}}
\put(220.0,430.0){\rule[-0.200pt]{4.818pt}{0.400pt}}
\put(198,430){\makebox(0,0)[r]{0.09}}
\put(846.0,430.0){\rule[-0.200pt]{4.818pt}{0.400pt}}
\put(220.0,482.0){\rule[-0.200pt]{4.818pt}{0.400pt}}
\put(198,482){\makebox(0,0)[r]{0.095}}
\put(846.0,482.0){\rule[-0.200pt]{4.818pt}{0.400pt}}
\put(220.0,535.0){\rule[-0.200pt]{4.818pt}{0.400pt}}
\put(198,535){\makebox(0,0)[r]{0.1}}
\put(846.0,535.0){\rule[-0.200pt]{4.818pt}{0.400pt}}
\put(220.0,113.0){\rule[-0.200pt]{0.400pt}{4.818pt}}
\put(220,68){\makebox(0,0){10}}
\put(220.0,515.0){\rule[-0.200pt]{0.400pt}{4.818pt}}
\put(328.0,113.0){\rule[-0.200pt]{0.400pt}{4.818pt}}
\put(328,68){\makebox(0,0){15}}
\put(328.0,515.0){\rule[-0.200pt]{0.400pt}{4.818pt}}
\put(435.0,113.0){\rule[-0.200pt]{0.400pt}{4.818pt}}
\put(435,68){\makebox(0,0){20}}
\put(435.0,515.0){\rule[-0.200pt]{0.400pt}{4.818pt}}
\put(543.0,113.0){\rule[-0.200pt]{0.400pt}{4.818pt}}
\put(543,68){\makebox(0,0){25}}
\put(543.0,515.0){\rule[-0.200pt]{0.400pt}{4.818pt}}
\put(651.0,113.0){\rule[-0.200pt]{0.400pt}{4.818pt}}
\put(651,68){\makebox(0,0){30}}
\put(651.0,515.0){\rule[-0.200pt]{0.400pt}{4.818pt}}
\put(758.0,113.0){\rule[-0.200pt]{0.400pt}{4.818pt}}
\put(758,68){\makebox(0,0){35}}
\put(758.0,515.0){\rule[-0.200pt]{0.400pt}{4.818pt}}
\put(866.0,113.0){\rule[-0.200pt]{0.400pt}{4.818pt}}
\put(866,68){\makebox(0,0){40}}
\put(866.0,515.0){\rule[-0.200pt]{0.400pt}{4.818pt}}
\put(220.0,113.0){\rule[-0.200pt]{155.621pt}{0.400pt}}
\put(866.0,113.0){\rule[-0.200pt]{0.400pt}{101.660pt}}
\put(220.0,535.0){\rule[-0.200pt]{155.621pt}{0.400pt}}
\put(45,324){\makebox(0,0){$C(i)$}}
\put(543,23){\makebox(0,0){$i$}}
\put(220.0,113.0){\rule[-0.200pt]{0.400pt}{101.660pt}}
\put(220,226){\raisebox{-.8pt}{\makebox(0,0){$\Diamond$}}}
\put(242,236){\raisebox{-.8pt}{\makebox(0,0){$\Diamond$}}}
\put(263,270){\raisebox{-.8pt}{\makebox(0,0){$\Diamond$}}}
\put(285,264){\raisebox{-.8pt}{\makebox(0,0){$\Diamond$}}}
\put(306,240){\raisebox{-.8pt}{\makebox(0,0){$\Diamond$}}}
\put(328,255){\raisebox{-.8pt}{\makebox(0,0){$\Diamond$}}}
\put(349,245){\raisebox{-.8pt}{\makebox(0,0){$\Diamond$}}}
\put(371,242){\raisebox{-.8pt}{\makebox(0,0){$\Diamond$}}}
\put(392,267){\raisebox{-.8pt}{\makebox(0,0){$\Diamond$}}}
\put(414,260){\raisebox{-.8pt}{\makebox(0,0){$\Diamond$}}}
\put(435,239){\raisebox{-.8pt}{\makebox(0,0){$\Diamond$}}}
\put(457,258){\raisebox{-.8pt}{\makebox(0,0){$\Diamond$}}}
\put(478,257){\raisebox{-.8pt}{\makebox(0,0){$\Diamond$}}}
\put(500,254){\raisebox{-.8pt}{\makebox(0,0){$\Diamond$}}}
\put(521,249){\raisebox{-.8pt}{\makebox(0,0){$\Diamond$}}}
\put(543,254){\raisebox{-.8pt}{\makebox(0,0){$\Diamond$}}}
\put(565,239){\raisebox{-.8pt}{\makebox(0,0){$\Diamond$}}}
\put(586,248){\raisebox{-.8pt}{\makebox(0,0){$\Diamond$}}}
\put(608,242){\raisebox{-.8pt}{\makebox(0,0){$\Diamond$}}}
\put(629,233){\raisebox{-.8pt}{\makebox(0,0){$\Diamond$}}}
\put(651,250){\raisebox{-.8pt}{\makebox(0,0){$\Diamond$}}}
\put(672,238){\raisebox{-.8pt}{\makebox(0,0){$\Diamond$}}}
\put(694,239){\raisebox{-.8pt}{\makebox(0,0){$\Diamond$}}}
\put(715,239){\raisebox{-.8pt}{\makebox(0,0){$\Diamond$}}}
\put(737,239){\raisebox{-.8pt}{\makebox(0,0){$\Diamond$}}}
\put(758,243){\raisebox{-.8pt}{\makebox(0,0){$\Diamond$}}}
\put(780,255){\raisebox{-.8pt}{\makebox(0,0){$\Diamond$}}}
\put(801,243){\raisebox{-.8pt}{\makebox(0,0){$\Diamond$}}}
\put(823,251){\raisebox{-.8pt}{\makebox(0,0){$\Diamond$}}}
\put(844,245){\raisebox{-.8pt}{\makebox(0,0){$\Diamond$}}}
\put(866,255){\raisebox{-.8pt}{\makebox(0,0){$\Diamond$}}}
\put(220.0,175.0){\rule[-0.200pt]{0.400pt}{24.331pt}}
\put(210.0,175.0){\rule[-0.200pt]{4.818pt}{0.400pt}}
\put(210.0,276.0){\rule[-0.200pt]{4.818pt}{0.400pt}}
\put(242.0,202.0){\rule[-0.200pt]{0.400pt}{16.622pt}}
\put(232.0,202.0){\rule[-0.200pt]{4.818pt}{0.400pt}}
\put(232.0,271.0){\rule[-0.200pt]{4.818pt}{0.400pt}}
\put(263.0,225.0){\rule[-0.200pt]{0.400pt}{21.681pt}}
\put(253.0,225.0){\rule[-0.200pt]{4.818pt}{0.400pt}}
\put(253.0,315.0){\rule[-0.200pt]{4.818pt}{0.400pt}}
\put(285.0,242.0){\rule[-0.200pt]{0.400pt}{10.359pt}}
\put(275.0,242.0){\rule[-0.200pt]{4.818pt}{0.400pt}}
\put(275.0,285.0){\rule[-0.200pt]{4.818pt}{0.400pt}}
\put(306.0,212.0){\rule[-0.200pt]{0.400pt}{13.490pt}}
\put(296.0,212.0){\rule[-0.200pt]{4.818pt}{0.400pt}}
\put(296.0,268.0){\rule[-0.200pt]{4.818pt}{0.400pt}}
\put(328.0,232.0){\rule[-0.200pt]{0.400pt}{11.081pt}}
\put(318.0,232.0){\rule[-0.200pt]{4.818pt}{0.400pt}}
\put(318.0,278.0){\rule[-0.200pt]{4.818pt}{0.400pt}}
\put(349.0,241.0){\rule[-0.200pt]{0.400pt}{1.686pt}}
\put(339.0,241.0){\rule[-0.200pt]{4.818pt}{0.400pt}}
\put(339.0,248.0){\rule[-0.200pt]{4.818pt}{0.400pt}}
\put(371.0,235.0){\rule[-0.200pt]{0.400pt}{3.373pt}}
\put(361.0,235.0){\rule[-0.200pt]{4.818pt}{0.400pt}}
\put(361.0,249.0){\rule[-0.200pt]{4.818pt}{0.400pt}}
\put(392.0,263.0){\rule[-0.200pt]{0.400pt}{1.927pt}}
\put(382.0,263.0){\rule[-0.200pt]{4.818pt}{0.400pt}}
\put(382.0,271.0){\rule[-0.200pt]{4.818pt}{0.400pt}}
\put(414.0,254.0){\rule[-0.200pt]{0.400pt}{2.650pt}}
\put(404.0,254.0){\rule[-0.200pt]{4.818pt}{0.400pt}}
\put(404.0,265.0){\rule[-0.200pt]{4.818pt}{0.400pt}}
\put(435.0,233.0){\rule[-0.200pt]{0.400pt}{2.650pt}}
\put(425.0,233.0){\rule[-0.200pt]{4.818pt}{0.400pt}}
\put(425.0,244.0){\rule[-0.200pt]{4.818pt}{0.400pt}}
\put(457.0,252.0){\rule[-0.200pt]{0.400pt}{2.891pt}}
\put(447.0,252.0){\rule[-0.200pt]{4.818pt}{0.400pt}}
\put(447.0,264.0){\rule[-0.200pt]{4.818pt}{0.400pt}}
\put(478.0,232.0){\rule[-0.200pt]{0.400pt}{12.286pt}}
\put(468.0,232.0){\rule[-0.200pt]{4.818pt}{0.400pt}}
\put(468.0,283.0){\rule[-0.200pt]{4.818pt}{0.400pt}}
\put(500.0,250.0){\rule[-0.200pt]{0.400pt}{1.927pt}}
\put(490.0,250.0){\rule[-0.200pt]{4.818pt}{0.400pt}}
\put(490.0,258.0){\rule[-0.200pt]{4.818pt}{0.400pt}}
\put(521.0,249.0){\usebox{\plotpoint}}
\put(511.0,249.0){\rule[-0.200pt]{4.818pt}{0.400pt}}
\put(511.0,250.0){\rule[-0.200pt]{4.818pt}{0.400pt}}
\put(543.0,245.0){\rule[-0.200pt]{0.400pt}{4.336pt}}
\put(533.0,245.0){\rule[-0.200pt]{4.818pt}{0.400pt}}
\put(533.0,263.0){\rule[-0.200pt]{4.818pt}{0.400pt}}
\put(565.0,234.0){\rule[-0.200pt]{0.400pt}{2.650pt}}
\put(555.0,234.0){\rule[-0.200pt]{4.818pt}{0.400pt}}
\put(555.0,245.0){\rule[-0.200pt]{4.818pt}{0.400pt}}
\put(586.0,237.0){\rule[-0.200pt]{0.400pt}{5.059pt}}
\put(576.0,237.0){\rule[-0.200pt]{4.818pt}{0.400pt}}
\put(576.0,258.0){\rule[-0.200pt]{4.818pt}{0.400pt}}
\put(608.0,240.0){\rule[-0.200pt]{0.400pt}{0.964pt}}
\put(598.0,240.0){\rule[-0.200pt]{4.818pt}{0.400pt}}
\put(598.0,244.0){\rule[-0.200pt]{4.818pt}{0.400pt}}
\put(629.0,231.0){\rule[-0.200pt]{0.400pt}{0.964pt}}
\put(619.0,231.0){\rule[-0.200pt]{4.818pt}{0.400pt}}
\put(619.0,235.0){\rule[-0.200pt]{4.818pt}{0.400pt}}
\put(651.0,245.0){\rule[-0.200pt]{0.400pt}{2.409pt}}
\put(641.0,245.0){\rule[-0.200pt]{4.818pt}{0.400pt}}
\put(641.0,255.0){\rule[-0.200pt]{4.818pt}{0.400pt}}
\put(672.0,237.0){\usebox{\plotpoint}}
\put(662.0,237.0){\rule[-0.200pt]{4.818pt}{0.400pt}}
\put(662.0,238.0){\rule[-0.200pt]{4.818pt}{0.400pt}}
\put(694.0,232.0){\rule[-0.200pt]{0.400pt}{3.373pt}}
\put(684.0,232.0){\rule[-0.200pt]{4.818pt}{0.400pt}}
\put(684.0,246.0){\rule[-0.200pt]{4.818pt}{0.400pt}}
\put(715.0,237.0){\rule[-0.200pt]{0.400pt}{1.204pt}}
\put(705.0,237.0){\rule[-0.200pt]{4.818pt}{0.400pt}}
\put(705.0,242.0){\rule[-0.200pt]{4.818pt}{0.400pt}}
\put(737.0,236.0){\rule[-0.200pt]{0.400pt}{1.204pt}}
\put(727.0,236.0){\rule[-0.200pt]{4.818pt}{0.400pt}}
\put(727.0,241.0){\rule[-0.200pt]{4.818pt}{0.400pt}}
\put(758.0,232.0){\rule[-0.200pt]{0.400pt}{5.300pt}}
\put(748.0,232.0){\rule[-0.200pt]{4.818pt}{0.400pt}}
\put(748.0,254.0){\rule[-0.200pt]{4.818pt}{0.400pt}}
\put(780.0,243.0){\rule[-0.200pt]{0.400pt}{5.541pt}}
\put(770.0,243.0){\rule[-0.200pt]{4.818pt}{0.400pt}}
\put(770.0,266.0){\rule[-0.200pt]{4.818pt}{0.400pt}}
\put(801.0,242.0){\rule[-0.200pt]{0.400pt}{0.723pt}}
\put(791.0,242.0){\rule[-0.200pt]{4.818pt}{0.400pt}}
\put(791.0,245.0){\rule[-0.200pt]{4.818pt}{0.400pt}}
\put(823.0,248.0){\rule[-0.200pt]{0.400pt}{1.686pt}}
\put(813.0,248.0){\rule[-0.200pt]{4.818pt}{0.400pt}}
\put(813.0,255.0){\rule[-0.200pt]{4.818pt}{0.400pt}}
\put(844.0,242.0){\rule[-0.200pt]{0.400pt}{1.204pt}}
\put(834.0,242.0){\rule[-0.200pt]{4.818pt}{0.400pt}}
\put(834.0,247.0){\rule[-0.200pt]{4.818pt}{0.400pt}}
\put(866.0,252.0){\rule[-0.200pt]{0.400pt}{1.445pt}}
\put(856.0,252.0){\rule[-0.200pt]{4.818pt}{0.400pt}}
\put(856.0,258.0){\rule[-0.200pt]{4.818pt}{0.400pt}}
\put(220,252){\usebox{\plotpoint}}
\put(220.00,252.00){\usebox{\plotpoint}}
\multiput(227,252)(20.756,0.000){0}{\usebox{\plotpoint}}
\multiput(233,252)(20.756,0.000){0}{\usebox{\plotpoint}}
\put(240.76,252.00){\usebox{\plotpoint}}
\multiput(246,252)(20.756,0.000){0}{\usebox{\plotpoint}}
\multiput(253,252)(20.756,0.000){0}{\usebox{\plotpoint}}
\put(261.51,252.00){\usebox{\plotpoint}}
\multiput(266,252)(20.756,0.000){0}{\usebox{\plotpoint}}
\multiput(272,252)(20.756,0.000){0}{\usebox{\plotpoint}}
\put(282.27,252.00){\usebox{\plotpoint}}
\multiput(285,252)(20.756,0.000){0}{\usebox{\plotpoint}}
\multiput(292,252)(20.756,0.000){0}{\usebox{\plotpoint}}
\put(303.02,252.00){\usebox{\plotpoint}}
\multiput(305,252)(20.756,0.000){0}{\usebox{\plotpoint}}
\multiput(311,252)(20.756,0.000){0}{\usebox{\plotpoint}}
\put(323.78,252.00){\usebox{\plotpoint}}
\multiput(324,252)(20.756,0.000){0}{\usebox{\plotpoint}}
\multiput(331,252)(20.756,0.000){0}{\usebox{\plotpoint}}
\multiput(337,252)(20.756,0.000){0}{\usebox{\plotpoint}}
\put(344.53,252.00){\usebox{\plotpoint}}
\multiput(351,252)(20.756,0.000){0}{\usebox{\plotpoint}}
\multiput(357,252)(20.756,0.000){0}{\usebox{\plotpoint}}
\put(365.29,252.00){\usebox{\plotpoint}}
\multiput(370,252)(20.756,0.000){0}{\usebox{\plotpoint}}
\multiput(377,252)(20.756,0.000){0}{\usebox{\plotpoint}}
\put(386.04,252.00){\usebox{\plotpoint}}
\multiput(390,252)(20.756,0.000){0}{\usebox{\plotpoint}}
\multiput(396,252)(20.756,0.000){0}{\usebox{\plotpoint}}
\put(406.80,252.00){\usebox{\plotpoint}}
\multiput(409,252)(20.756,0.000){0}{\usebox{\plotpoint}}
\multiput(416,252)(20.756,0.000){0}{\usebox{\plotpoint}}
\put(427.55,252.00){\usebox{\plotpoint}}
\multiput(429,252)(20.756,0.000){0}{\usebox{\plotpoint}}
\multiput(435,252)(20.756,0.000){0}{\usebox{\plotpoint}}
\multiput(442,252)(20.756,0.000){0}{\usebox{\plotpoint}}
\put(448.31,252.00){\usebox{\plotpoint}}
\multiput(455,252)(20.756,0.000){0}{\usebox{\plotpoint}}
\multiput(461,252)(20.756,0.000){0}{\usebox{\plotpoint}}
\put(469.07,252.00){\usebox{\plotpoint}}
\multiput(474,252)(20.756,0.000){0}{\usebox{\plotpoint}}
\multiput(481,252)(20.756,0.000){0}{\usebox{\plotpoint}}
\put(489.82,252.00){\usebox{\plotpoint}}
\multiput(494,252)(20.756,0.000){0}{\usebox{\plotpoint}}
\multiput(501,252)(20.756,0.000){0}{\usebox{\plotpoint}}
\put(510.58,252.00){\usebox{\plotpoint}}
\multiput(514,252)(20.756,0.000){0}{\usebox{\plotpoint}}
\multiput(520,252)(20.756,0.000){0}{\usebox{\plotpoint}}
\put(531.33,252.00){\usebox{\plotpoint}}
\multiput(533,252)(20.756,0.000){0}{\usebox{\plotpoint}}
\multiput(540,252)(20.756,0.000){0}{\usebox{\plotpoint}}
\put(552.09,252.00){\usebox{\plotpoint}}
\multiput(553,252)(20.756,0.000){0}{\usebox{\plotpoint}}
\multiput(559,252)(20.756,0.000){0}{\usebox{\plotpoint}}
\multiput(566,252)(20.756,0.000){0}{\usebox{\plotpoint}}
\put(572.84,252.00){\usebox{\plotpoint}}
\multiput(579,252)(20.756,0.000){0}{\usebox{\plotpoint}}
\multiput(585,252)(20.756,0.000){0}{\usebox{\plotpoint}}
\put(593.60,252.00){\usebox{\plotpoint}}
\multiput(598,252)(20.756,0.000){0}{\usebox{\plotpoint}}
\multiput(605,252)(20.756,0.000){0}{\usebox{\plotpoint}}
\put(614.35,252.00){\usebox{\plotpoint}}
\multiput(618,252)(20.756,0.000){0}{\usebox{\plotpoint}}
\multiput(625,252)(20.756,0.000){0}{\usebox{\plotpoint}}
\put(635.11,252.00){\usebox{\plotpoint}}
\multiput(638,252)(20.756,0.000){0}{\usebox{\plotpoint}}
\multiput(644,252)(20.756,0.000){0}{\usebox{\plotpoint}}
\put(655.87,252.00){\usebox{\plotpoint}}
\multiput(657,252)(20.756,0.000){0}{\usebox{\plotpoint}}
\multiput(664,252)(20.756,0.000){0}{\usebox{\plotpoint}}
\put(676.62,252.00){\usebox{\plotpoint}}
\multiput(677,252)(20.756,0.000){0}{\usebox{\plotpoint}}
\multiput(683,252)(20.756,0.000){0}{\usebox{\plotpoint}}
\multiput(690,252)(20.756,0.000){0}{\usebox{\plotpoint}}
\put(697.38,252.00){\usebox{\plotpoint}}
\multiput(703,252)(20.756,0.000){0}{\usebox{\plotpoint}}
\multiput(709,252)(20.756,0.000){0}{\usebox{\plotpoint}}
\put(718.13,252.00){\usebox{\plotpoint}}
\multiput(722,252)(20.756,0.000){0}{\usebox{\plotpoint}}
\multiput(729,252)(20.756,0.000){0}{\usebox{\plotpoint}}
\put(738.89,252.00){\usebox{\plotpoint}}
\multiput(742,252)(20.756,0.000){0}{\usebox{\plotpoint}}
\multiput(749,252)(20.756,0.000){0}{\usebox{\plotpoint}}
\put(759.64,252.00){\usebox{\plotpoint}}
\multiput(762,252)(20.756,0.000){0}{\usebox{\plotpoint}}
\multiput(768,252)(20.756,0.000){0}{\usebox{\plotpoint}}
\put(780.40,252.00){\usebox{\plotpoint}}
\multiput(781,252)(20.756,0.000){0}{\usebox{\plotpoint}}
\multiput(788,252)(20.756,0.000){0}{\usebox{\plotpoint}}
\multiput(794,252)(20.756,0.000){0}{\usebox{\plotpoint}}
\put(801.15,252.00){\usebox{\plotpoint}}
\multiput(807,252)(20.756,0.000){0}{\usebox{\plotpoint}}
\multiput(814,252)(20.756,0.000){0}{\usebox{\plotpoint}}
\put(821.91,252.00){\usebox{\plotpoint}}
\multiput(827,252)(20.756,0.000){0}{\usebox{\plotpoint}}
\multiput(833,252)(20.756,0.000){0}{\usebox{\plotpoint}}
\put(842.66,252.00){\usebox{\plotpoint}}
\multiput(846,252)(20.756,0.000){0}{\usebox{\plotpoint}}
\multiput(853,252)(20.756,0.000){0}{\usebox{\plotpoint}}
\put(863.42,252.00){\usebox{\plotpoint}}
\put(866,252){\usebox{\plotpoint}}
\vspace{-10mm}
\caption{$C(i)$ as a function of $i$
for $N_f=4$, $\beta=5.2500$ and $V=6^4$ 
using Eq. (6).}
\label{fig6}
\end{figure}

\end{document}